\documentclass[prb,twocolumn,showpacs,floatfix,amsmath,amssymb,superscriptaddress]{revtex4-1}

\usepackage{amsfonts}
\usepackage{stmaryrd}
\usepackage{bbm}
\usepackage{mathrsfs}
\usepackage{tipa}
\usepackage{amssymb}
\usepackage{txfonts}
\usepackage{graphicx}
\usepackage{dcolumn}
\usepackage{epstopdf}
\usepackage[colorlinks,linkcolor=blue,urlcolor=blue,citecolor=blue]{hyperref}
\usepackage{multirow}
\usepackage{subfigure}
\usepackage{url}
\usepackage{setspace}

\begin{document}

\title{Strong nonlinear optical response and transient symmetry switching in Type-\uppercase\expandafter{\romannumeral2} Weyl semimetal $\beta$-WP$_2$ }
\author{T. C. Hu}
\affiliation{International Center for Quantum Materials, School of Physics, Peking University, Beijing 100871, China}

\author{B. Su}
\affiliation{Beijing National Laboratory for Condensed Matter Physics, Institute of Physics, Chinese Academy of Sciences, 100190 Beijing, China}

\author{L. Y. Shi}
\affiliation{International Center for Quantum Materials, School of Physics, Peking University, Beijing 100871, China}

\author{Z. X. Wang}
\affiliation{International Center for Quantum Materials, School of Physics, Peking University, Beijing 100871, China}

\author{L. Yue}
\affiliation{International Center for Quantum Materials, School of Physics, Peking University, Beijing 100871, China}

\author{S. X. Xu}
\affiliation{International Center for Quantum Materials, School of Physics, Peking University, Beijing 100871, China}

\author{S. J. Zhang}
\affiliation{International Center for Quantum Materials, School of Physics, Peking University, Beijing 100871, China}

\author{Q. M. Liu}
\affiliation{International Center for Quantum Materials, School of Physics, Peking University, Beijing 100871, China}

\author{Q. Wu}
\affiliation{International Center for Quantum Materials, School of Physics, Peking University, Beijing 100871, China}

\author{R. S. Li}
\affiliation{International Center for Quantum Materials, School of Physics, Peking University, Beijing 100871, China}

\author{X. Y. Zhou}
\affiliation{International Center for Quantum Materials, School of Physics, Peking University, Beijing 100871, China}

\author{J. Y. Yuan}
\affiliation{International Center for Quantum Materials, School of Physics, Peking University, Beijing 100871, China}

\author{D. Wu}
\affiliation{Beijing Academy of Quantum Information Sciences, Beijing 100913, China}

\author{Z. G. Chen}
\affiliation{Beijing National Laboratory for Condensed Matter Physics, Institute of Physics, Chinese Academy of Sciences, 100190 Beijing, China}
\affiliation{Songshan Lake Materials Laboratory, Dongguan, Guangdong 523808, China}

\author{T. Dong}
\affiliation{International Center for Quantum Materials, School of Physics, Peking University, Beijing 100871, China}

\author{N. L. Wang}
\email{nlwang@pku.edu.cn}
\affiliation{International Center for Quantum Materials, School of Physics, Peking University, Beijing 100871, China}
\affiliation{Beijing Academy of Quantum Information Sciences, Beijing 100913, China}

\begin{abstract}

 The topological Weyl semimetals with peculiar band structure exhibit novel nonlinear optical enhancement phenomena even for light at optical wavelengths. While many intriguing nonlinear optical effects were constantly uncovered in type-\uppercase\expandafter{\romannumeral1} semimetals, few experimental works focused on basic nonlinear optical properties in type-\uppercase\expandafter{\romannumeral2} Weyl semimetals. Here we perform a fundamental static and time-resolved second harmonic generation (SHG) on the three dimensional Type-\uppercase\expandafter{\romannumeral2} Weyl semimetal candidate $\beta$-WP$_2$. Although $\beta$-WP$_2$ exhibits extremely high conductivity and an extraordinarily large mean free path, the second harmonic generation is unscreened by conduction electrons, we observed rather strong SHG response compared to non-topological polar metals and archetypal ferroelectric insulators. Additionally, our time-resolved SHG experiment traces ultrafast symmetry switching and reveals that polar metal $\beta$-WP$_2$ tends to form inversion symmetric metastable state after photo-excitation. Intense femtosecond laser pulse could optically drive symmetry switching and tune nonlinear optical response on ultrafast timescales although the interlayer coupling of $\beta$-WP$_2$ is very strong. Our work is illuminating for the polar metal nonlinear optics and potential ultrafast topological optoelectronic applications.
\end{abstract}

\pacs{Valid PACS appear here}

\maketitle

\section{INTRODUCTION}
  
  As topological quantum materials, Weyl semimetals (WSMs) exhibit a unique linear band dispersion in close vicinity of the Fermi energy described by the Weyl equation\cite{Weyl1929}. Pairing Weyl nodes with left and right chirality occur in the crossing of conduction and valence bands\cite{Xu2016,Yao2019}. These nodes act as monopoles of Berry curvature in momentum space. The emergent low-energy quasiparticle excitations in WSMs are regarded as massless fermions\cite{Xu2015,Lv2015}. WSMs with the quasiparticles require the breaking of at least space inversion or time-reversal symmetry\cite{PhysRevX.5.011029,Xu2020}. Compare to type-\uppercase\expandafter{\romannumeral1} WSMs, the Lorentz invariance is violated and the Weyl cone is tilted in the momentum space in type-\uppercase\expandafter{\romannumeral2} WSMs\cite{Soluyanov2015,Armitage2018}. The novel topological band structure can be manifested in intriguing electronic and optical properties in condensed-matter physics\cite{Ali2014,Kumar2017,Ma2019}. 

  Recently, WSMs have aroused tremendous interests for their spectacular nonlinear optical properties and future potential technological applications\cite{Chi2018,Liu2020,Zhang2020}. Some  nonlinear optical phenomena such as the giant second harmonic generation, colossal mid-infrared bulk photovoltaic effect and chiral Terahertz wave emission has been observed in type-\uppercase\expandafter{\romannumeral1} WSM TaAs\cite{Wu2017,Osterhoudt2019,Sirica2019,Gao2020}. While it was proposed that type-\uppercase\expandafter{\romannumeral2} Weyl fermions could emerge from nonmagnetic layered compound T$_d$ phase WTe$_2$/MoTe$_2$\cite{Soluyanov2015,Sun2015} and three dimensional compound $\beta$-WP$_2$/MoP$_2$\cite{Autes2016}. Minority works mainly focus on their 
 primary nonlinear optical properties\cite{Drueke2021,Lv2021}. SHG is a fundamental nonlinear optical process and a sensitive probe for tracing crystal symmetry switching\cite{Hsieh2011,Zhao2016,Harter2017}. Electric-dipole (ED) allowed SHG response is expected to be dominant in the bulk crystal with lacking inversion symmetry structure and the SHG intensity reflects the magnitude of second-order susceptibility $\chi^{(2)}$, defined by the relation P$^{(2)}_i$=$\epsilon_{0}\chi_{ijk}E{_j}E{_k}$ \cite{Fiebig2005}. The magnitude of $\chi^{(2)}$ is significant in nonlinear optics and close connected to some other nonlinear optical processes such as optical rectification and photocurrents generating.

\begin{figure*}[htbp]
	\centering
	\includegraphics[width=18.5cm]{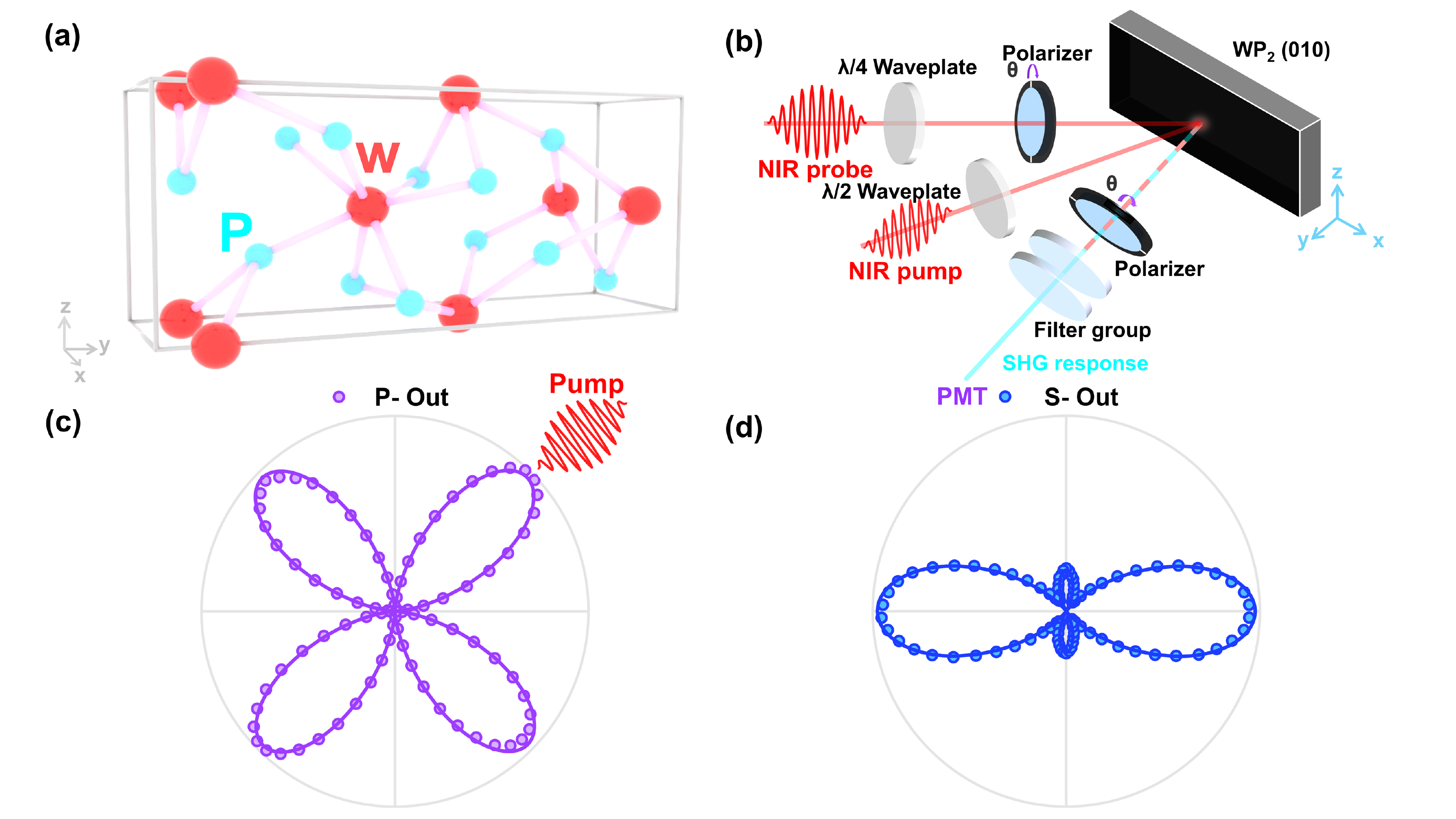}\\
	\caption{\textbf{Sample structure and characterizations for $\beta$-WP$_2$.} (a) Crystal structure of $\beta$-WP$_2$ (b) Schematic of the reflection static/time-resolved SHG spectroscopy geometry, focus lens are omitted. Polarization SHG scans (c) for the parallel-polarized (P-Out) and (d) for the perpendicular-polarized SH photons (S-Out) relative to the scattering plane while the incidence wired-polarizer is rotated. The solid curves are theory fit results to C$_{2v}$  point group symmetry. The 
polarization of incidence is kept 45° and the polarization of outgoing is kept parallel as depicted in (c) with  45° incidence for fluence dependent TR-SHG measurement.
}\label{Fig:1}
\end{figure*}

  In this work, we report an experimental study of static and time-resolved SHG in the three dimensional type-\uppercase\expandafter{\romannumeral2} WSM candidate bulk $\beta$-WP$_2$ single crystal. In static SHG measurement, we verify the C$_{2v}$ point group of $\beta$-WP$_2$ and estimate the size of the second-order susceptibility $\chi^{(2)}$. Although $\beta$-WP$_2$ exhibits extremely good metallic properties with large mean free path, we observe rather strong SHG response at near-infrared 800 nm wavelength excitation and the largest nonlinear susceptibility $|\chi_{xxz}^{(2)}|$ of WP$_2$ is as the same level of the widely-used electro-optics crystal GaAs\cite{Bergfeld2003} and ZnTe\cite{PhysRevB.58.10494}. Additionally, we find that light can non-thermally drive novel transient symmetry switching on 
 hundreds of picoseconds timescales. The maximum SHG response decreases at about 70$\%$ with 4 mJ/cm$^2$ laser pump fluence. It indicates that $\beta$-WP$_2$ has tendency to switch to the inversion symmetric (topological trivial) metastable state and the topological nonlinear optical properties could be tuned by light. The transient large SHG response comes from the lattice distortion which is tied to topology. 
 The ultrafast switching is accompanied by a distinct softening of the  A$_1$ phonon. This behaviour is generally related to the structural instability. In previous work, light-driven interlayer shear displacements plays an essential role in two dimensional layered type-\uppercase\expandafter{\romannumeral2} WSMs T$_d$-XTe$_2$ (X=W,Mo) ultrafast symmetry switching. They could be induced a structural phase transition from T$_d$ to 1T' phase and completely to the topological trivial metastable state by light\cite{Sie2019,Zhang2019}. For three dimensional compound $\beta$-WP$_2$, since the strength of interlayer coupling is very strong and robust, ultrafast bond softening mechanism driving by light or other mechanism should be considered. The results have implications for optical control of the symmetry and topological properties of the transition metal phosphides and related materials. 

\begin{figure*}[htbp]
	\centering
	\includegraphics[width=17.5cm]{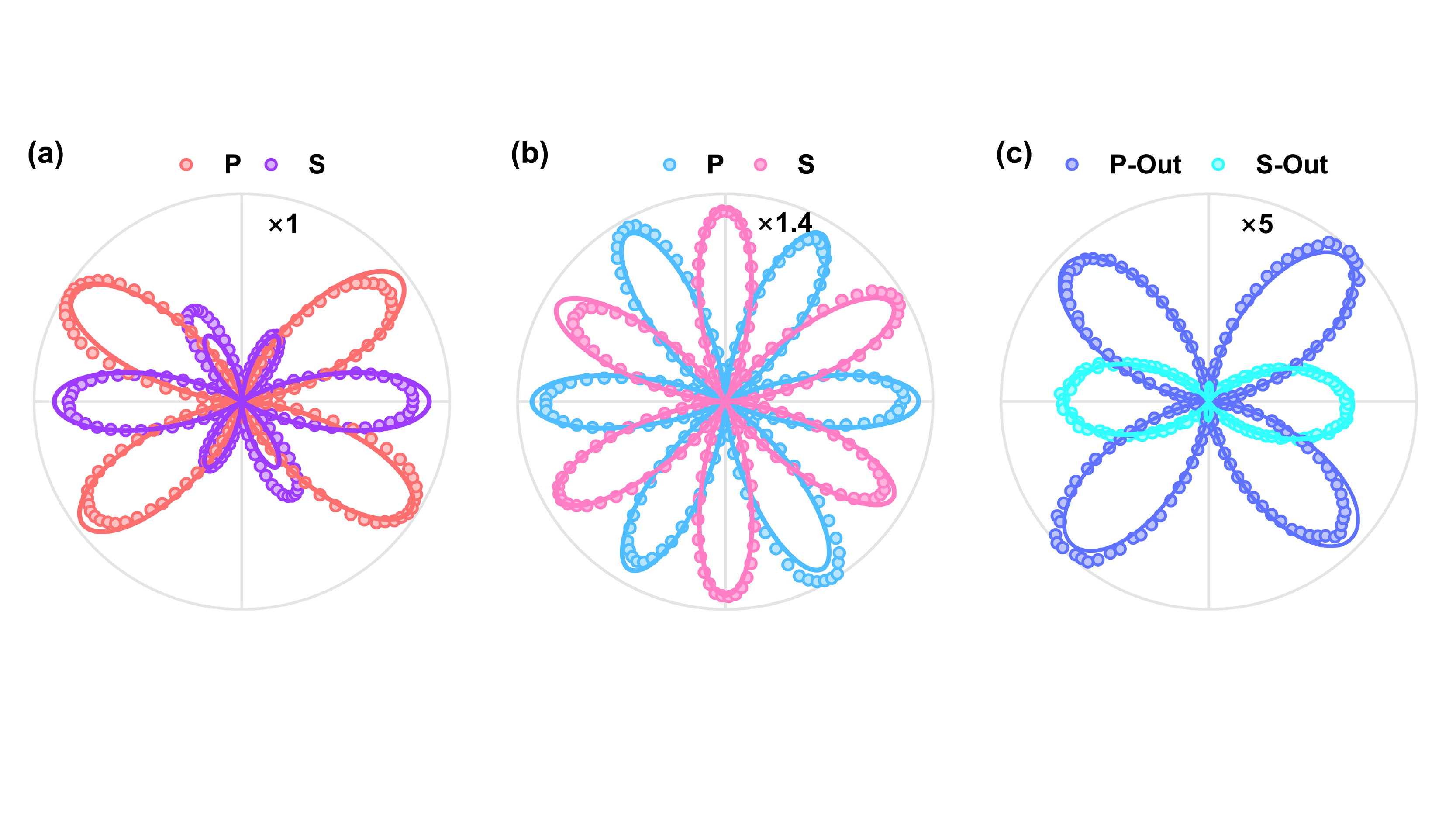}\\
	\caption{\textbf{Benchmark static SHG measurements.} Benchmark 11° incident 800 nm SHG experiments on (a) (110) ZnTe, (b) (111) intrinsic insulating GaAs wafer ($\rho$ $>$1$\times$10$^7$   $\Omega\cdot$cm) and (c) (010) $\beta$-WP$_2$, the signal is normalized to the maximum SHG intensity of ZnTe.
}\label{Fig:2}
\end{figure*}

\section{RESULTS AND DISCUSSION}
$\beta$-WP$_2$ is a three dimensional compound and typically needle-shaped with shiny luster (Optical micrograph of $\beta$-WP$_2$ is shown in Fig.S1.). Previous Raman experimental results indirectly indicate that $\beta$-WP$_2$ crystallizes into a non-centrosymmetric orthorhombic Cmc2$_1$ (No.36) space group\cite{Su2019,Wulferding2020,Osterhoudt2021}. A simple illustration of the crystal structure is shown in Figure \ref{Fig:1}(a). It contains a twofold screw axis along the crystalline z-axis, a glide plane perpendicular to the y-axis, and a mirror plane perpendicular to the x-axis. Compare to the first predicted type-\uppercase\expandafter{\romannumeral2} WSMs candidate layered T$_d$-WTe$_2$/MoTe$_2$, the interlayer coupling of $\beta$-WP$_2$ are much stronger while crystal point group is the same as C$_{2v}$. No structural or other phase transition has been observed in $\beta$-WP$_2$ at 4-300 K range yet\cite{PhysRevB.96.121107,Wulferding2020,Su2019}.

Since $\beta$-WP$_2$ is a nonmagnetic compound, inversion symmetry breaking is prerequisite for topological nontrivial band structure. It is expected to produce fundamental SHG response from ED contribution. Our model reflecting staic/time-resolved SHG optical measurement setup is depicted in Figure \ref{Fig:1}(b). In our setup, the WP$_2$ sample is kept static while the polarization of incidence or outgoing beam is rotated at least. The long x-axis of WP$_2$ xz plane is attached on the sample holder along the parallel direction. The electric field polarizations of the obliquely incident fundamental beam (in) and outgoing SHG beam (out) can be independently selected to lie either parallel or perpendicular to the crystal high-symmetry axis. The SHG response data are acquired by measuring the parallel(P-out) and perpendicular (S-out) component of SHG intensity reflected from the (010) surface of WP$_2$ as a function of the polarization at 45° incidence beam. In crystals with C$_{2v}$ point group symmetry,  five independent non-vanishing elements of $\chi^{(2)}$, there are  $\chi^{(2)}_{xxz}$=$\chi^{(2)}_{xzx}$, $\chi^{(2)}_{yyz}$=$\chi^{(2)}_{yzy}$, $\chi^{(2)}_{zxx}$, $\chi^{(2)}_{zyy}$, $\chi^{(2)}_{zzz}$. Hence rotating oblique 45° xz plane fundamental incidence polarization could simultaneously detect ED-SHG responses from all the symmetry-allowed tensor elements. The fit lobes for P-out and S-out configurations are $I_{P-out}(\theta)$$\propto$$\frac{1}{4}$${|\chi_{xxz}+\chi_{yyz}|}^2$$sin^2$$(2\theta)$, $I_{S-out}(\theta)$
$\propto$${|\frac{1}{2}(\chi_{zxx}+\chi_{zyy})cos^2(\theta)+\chi_{zzz}sin^2(\theta)|}^2$, respectively, where the polarization of incidence rotates at an angle $\theta$. The results of the two scan patterns are shown in Figure \ref{Fig:1}(c) and \ref{Fig:1}(d). All the solid fit curves for SHG responses derived from the C$_{2v}$ point group. Detailed formula derivation procedure for SHG patterns can be found in supplementary information. Our static rotation SHG results offer a very direct and explicit evidence that the point group symmetry of the $\beta$-WP$_2$ is C$_{2v}$.

The intensity of ED-SHG response reflect the magnitude of second-order nonlinear susceptibility. And the order of magnitude of $\chi^{(2)}$ is significative and generally related to some other  nonlinear optical processes such as optical rectification, difference frequency generation and generating photocurrents. Our optical setup and measuring scheme for the estimation of $\chi^{(2)}$ refers to previous work in TaAs\cite{Wu2017}. Since it is liable to distinguish each tensor component $\chi^{(2)}$ contribution, the angle of incidence is set as near-normalized (11°). We used insulating (111) GaAs wafer ($\rho>$1$\times$10$^7$ $\Omega\cdot$cm) and 1mm-thick (110) ZnTe archetypal electro-optic crystals as benchmarks. Both benchmark crystals have large and well-characterized SHG responses, only one non-vanishing tensor $\chi_{xyz}^{(2)}$ at fundamental 800 nm wavelength ($\hbar\omega$=1.55 eV) for their high symmetry point group (T$_d$). The SHG probe fluence is kept with 0.30 mJ/cm$^2$ for the measurements. Surprisingly, the SHG response of $\beta$-WP$_2$ is rather strong that unscreened by a number of conduction electrons. As seen in the Fig.\ref{Fig:2}(a), \ref{Fig:2}(b) and \ref{Fig:2}(c), the signal intensity is normalized to the peak SHG intensity in ZnTe. The peak intensity formulas are derived in Supporting information. In previous literature, the experimental refractive index of bulk GaAs, ZnTe and $\beta$-WP$_2$ can be found. Taking $|\chi_{xyz}^{(2)}|$ value of ZnTe as reference, by Bloembergan and Pershan expression\cite{Bloembergen1962}, we could calculate the maximum  of $\beta$-WP$_2$ is about 150 pV/m. The order of the estimated $|\chi_{xxz}^{(2)}|$ is approximate to the theoretical calculation peak value of $|\chi^{(2)}|$ 
for representative ferroelectric insulators BiFeO$_3$ ($|\chi_{zzz}^{(2)}|$= 260 pV/m, fundamental wavelength 500 nm) and BaTiO$_3$ ($|\chi_{zzz}^{(2)}|$= 200 pV/m, fundamental wavelength 170 nm)\cite{Ju2009,Young2012}, even one or two order larger than LiNbO$_3$, LiTaO$_3$ and topological trivial polar metal LiOsO$_3$ at 800 nm fundamental wavelength \cite{Padmanabhan2018}.

\begin{figure*}[htbp]
	\centering
	\includegraphics[width=17.5cm]{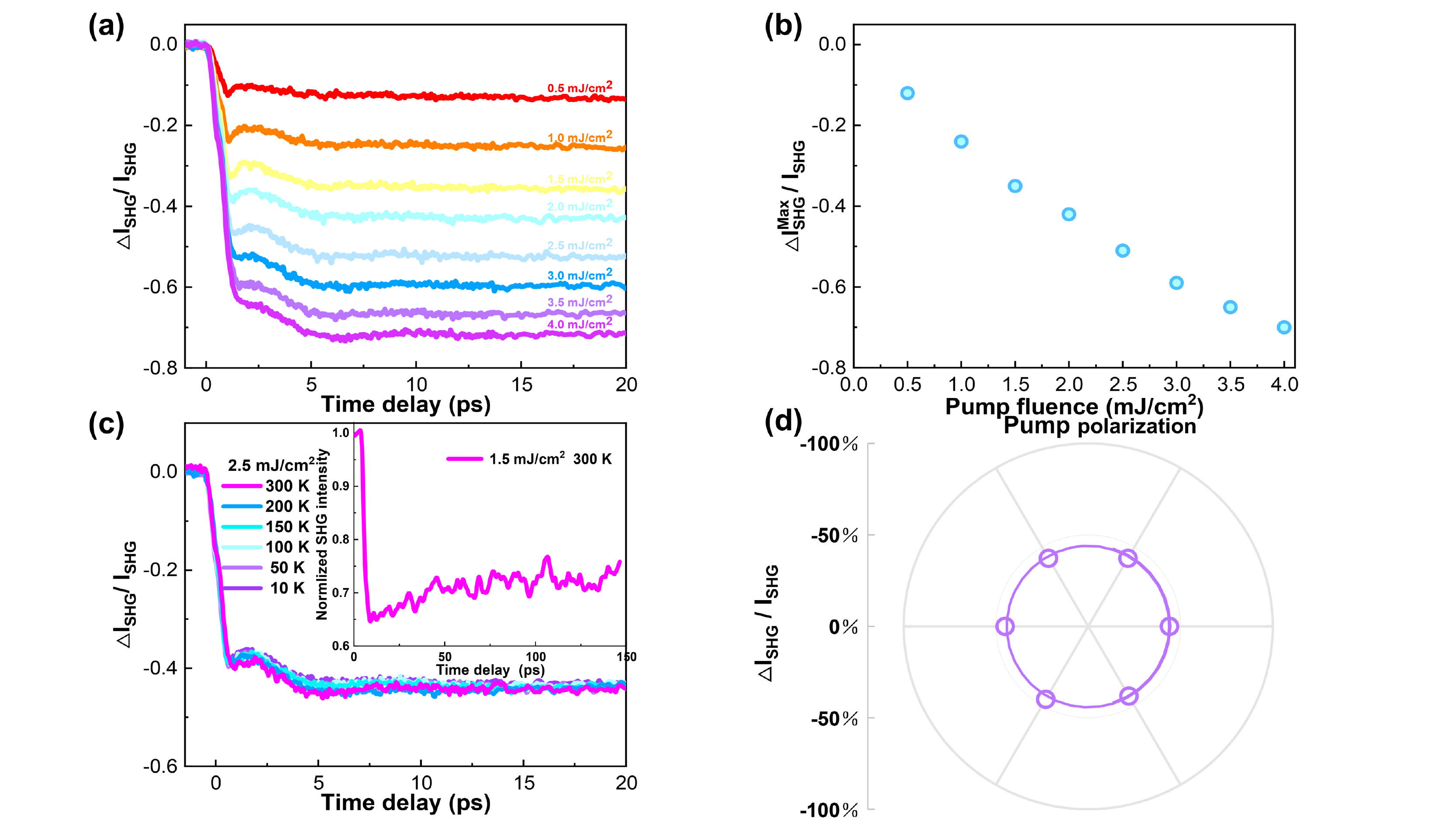}\\
	\caption{\textbf{Time-resolved SHG responses of $\beta$-WP$_2$.} (a) Pump-induced SHG time traces of WP$_2$ at various pump fluences. (b) SHG intensity changes at increasing fluences of 1.55 eV photon energy. (c) SHG time traces of WP$_2$ with 2.5 mJ/cm$^2$ fluence at different temperature points. Inset: The SHG intensity change with 1.5mJ/cm$^2$ for much longer timescales. (d) The nearly isotropic SHG reduction response in pump polarization with 2 mJ/cm$^2$ fluence.
}\label{Fig:3}
\end{figure*}

The magnitude of ED-SHG response depends on several possible factors. Generally, it is recognized that the intensity of SHG is close-linked to the degree of the crystal polarization. So insulating ferroelectric is more ideal for strong ED-SHG response in a sense. Appearance of conduction electrons in a crystal can screen and weaken the polarization and SHG responses\cite{Anderson1965}. Since $\beta$-WP$_2$ exhibits extremely high conductivity and an extraordinarily large mean free path, it is unexpected to observe strong SHG signal. Meanwhile, strong SHG response was hardly observed in topological trivial polar metal such as LiOsO$_3$\cite{Padmanabhan2018}. Obviously, the conventional displacive ferroelectric polarization perspective inadequately illuminates the strong SHG response from topological polar metal. The strong SHG response may arise from the peculiar band topology contributions in theoretical model proposed by Morimoto and Nagaosa. This theory deduced concise expression with explicit geometrical meaning for $\chi^{(2)}$  and its possible relation to the existence of Weyl nodes. However, the energy of fundamental excitation at optical wavelength is away from the low energies Weyl physics. Recently, resonance-enhanced effect from other bands was proposed and consist with their first-principle calculation\cite{Patankar2018,Lu2022}. Previous infrared spectroscopy studies indicate much higher carrier densities in $\beta$-WP$_2$\cite{Xu2016-,Su2019}, thus it is reasonable to expect that the SHG response is weakened in view of carriers screening effect in ferroelectric polarization perspective when compared with type-\uppercase\expandafter{\romannumeral1} WSM TaAs. The origin of strong SHG response in polar WSMs is still controversial, topological band contributions, band resonance-enhanced and carriers screening effect are all possible influences for the SHG response\cite{Wu2017,Patankar2018,Li2018,Lu2022}. In $\beta$-WP$_2$, the weyl nodes are located hundreds of meV below fermi energy\cite{Autes2016,Yao2019}. The relative large energy scale provides a better platform for studying the strong topological nonlinear optical response. Since the value of $\chi^{(2)}$ is dependent on the frequency, more theoretical and calculation work should be stimulated for the attractive strong topological nonlinear optical responses in type-\uppercase\expandafter{\romannumeral1} and type-\uppercase\expandafter{\romannumeral2} polar WSMs based on band structure and wavefunctions.

\begin{figure*}[htbp]
	\centering
	\includegraphics[width=17.5cm]{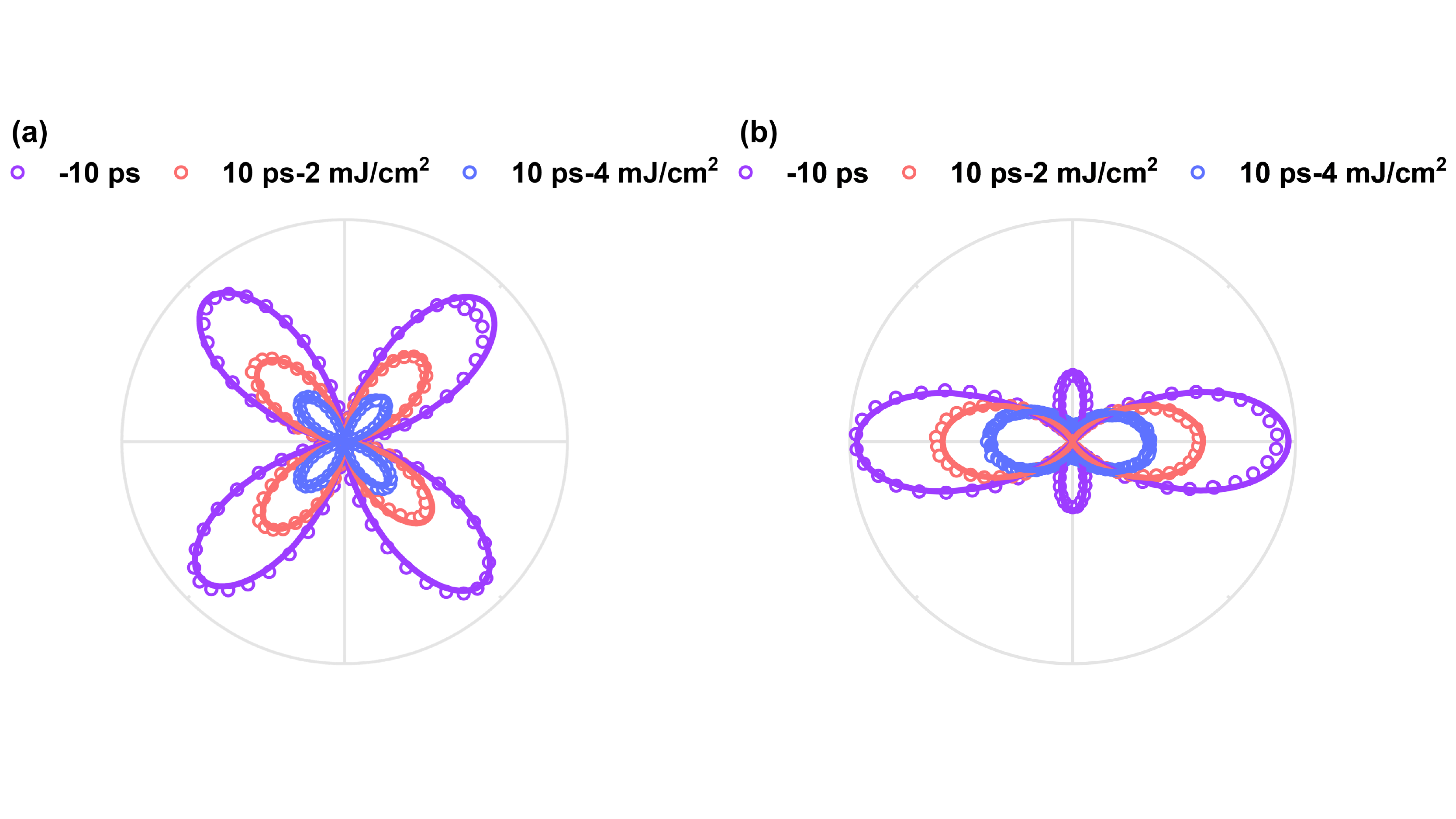}\\
	\caption{\textbf{Snapshots of transient SHG patterns.} With a linearly polarized pump excitation, the SHG patterns for S/P-Out configurations with 45° probe incidence for various delay time and pump fluence. Fits of the pattern assuming the crystal point symmetries of C$_{2v}$ are shown as solid.  
}\label{Fig:4}
\end{figure*}

Symmetry plays a central role in conventional and topological phases of matter, intense femtosecond laser pulse can optically drive transient symmetry switching and make novel metastable or hidden states in some quantum materials\cite{Stojchevska2014,Wang2021,Bao2022,Sirica2022}. Electric dipole allowed SHG arises from a non-zero second-order susceptibility $\chi^{(2)}$, as shown in non-centrosymmetric topological systems\cite{Morimoto2016,Wu2017,Drueke2021}. Thus, time-resolved SHG measurement is a very sensitive technique for monitoring ultrafast transient symmetry switching processes in 
these topological materials. Our optical setup for tr-SHG measurements is shown in \ref{Fig:1}(b). The pump pulse is normal to the xz plane, the polarization of incidence and reflecting probe pulse is set to make the static SHG response at maximum value, with 45° incidence P-out configuration. (The polarization direction of incidence is depicted in Fig.\ref{Fig:1}(c).)

After the femtosecond pump pulse arrives ($\Delta$t$\approx$1.5 ps), the SHG intensity decreases saliently. Figure.\ref{Fig:3}(a) shows the measured SHG time traces from WP$_2$ at various pump fluences. At rather low fluence excitation (about 0.5 mJ/cm$^2$, red curve), the time dependent SHG intensity shows remarkable reduction. With 4.0 mJ/cm$^2$ pump fluence, the SHG intensity plummets drastically, about 70$\%$ reduction. Continue to increase pump fluence until the damage threshold, the value of $\Delta$I$_{max}$/I keeps almost unchanged. The much longer timescales intensity variation is shown in inset of Fig.\ref{Fig:3}(c). The metastable state persists with more than one hundred picoseconds and slowly relax to the initial state before the next pump pulse arrives. Figure.\ref{Fig:3}(c) shows the temperature-dependent relative change of the SHG intensity at fixed pump fluence about 2.5 mJ/cm$^2$ as a function of the delay time. Rather slight maximum SHG intensity reduction difference of temperature variations in Fig.\ref{Fig:3}(c) and no high temperature structural transition was observed, manifesting non-thermal symmetry switching process. Varying the linear polarization of the excitation pulse in the xz plane relative to the x-axis as shown in Fig.\ref{Fig:3}(d), no distinct amplitude change could be found in $\Delta$I$_{max}$/I. The results suggest the tr-SHG responses are almost isotropic in the pump polarization. Compare to the previous work in layered T$_d$-WTe$_2$, light with sufficient electric field strength can drive it to a completely topological trivial state, indicating by almost no SHG response\cite{Sie2019}. Firstly, one possible reason is that the Gaussian-like pump beam profile is nonuniform, evidently much larger intensity in the center of the beam than at the edge region. Secondly, the bonding state in WP$_2$ is very strong rather than weak van del Waals coupling. For this reason, it is very tough to drive the atoms move to the totally centrosymmetric positions with SHG response decreasing to almost undetected. In general, the lattice symmetry is dominant in ED-SHG response that it is often taken as a sensitive tool to detect the structural symmetry change of a compound such as GaAs\cite{Saeta1991}. It is appropriate to take the structural change into account for the transient symmetry switching.

\begin{figure*}[htbp]
	\centering
	\includegraphics[width=17.5cm]{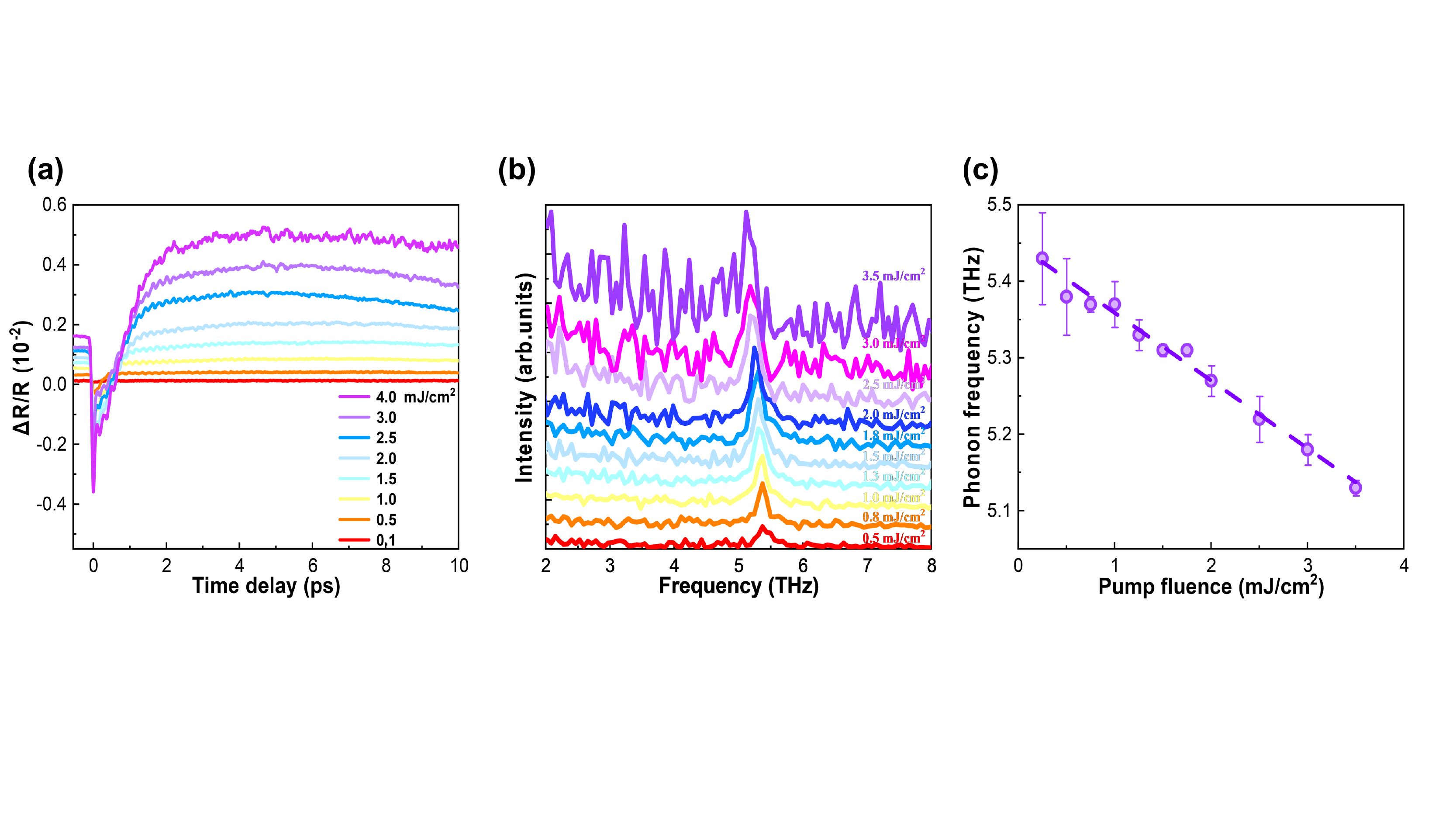}\\
	\caption{\textbf{The fluence-dependent coherent vibrational dynamics.} (a) The transient pump-induced reflectivity at 800 nm with 0.1-4 mJ/cm$^2$ fluence. 
	(b) Fourier transformation of the oscillation after subtracting the background.
	(c) The pump fluence dependence of A$_1$ phonon mode frequency.
}\label{Fig:5}
\end{figure*}

Following near-infrared (1.55 eV) light excitation, figure.\ref{Fig:4}(a) and \ref{Fig:4}(b) show pronounced SHG signal changes in the two polarization scan lobes before/after pump pulse arriving. In the P/S-out SHG scan lobes, the SHG response is still consistent with the equilibrium C$_{2v}$ point group symmetry fit at 10 ps, but with much less SHG responses. The maxmium SHG tensor element $|\chi_{xxz}|$ shows a reduction of nearly 80$\%$. Considering the pump electric field polarization and perturbations are in the xz-plane, the results suggest a connection between the polar axis and the strong optical nonlinearity of the $\beta$-WP$_2$. The snapshots of SHG polarization patterns indicate the specific reversible ultrafast symmetry switching and a strong tendency to the inversion symmetric state. Thus, intense laser pulse could manipulate the nonlinear optical properties drastically in WP$_2$ in an ultrafast fashion.

Due to rather weak van der Waals bonding, interlayer shear displacements could be driven by intense light in T$_d$-XTe$_2$(W,Mo). The intensity of low-frequency shear mode phonon would recede in time-resolved reflectivity experiment accompanied by the light-induced structural transition. However, previous Raman results of $\beta$-WP$_2$ point out that no such a low-frequency shear motion phonon existed. In order to get further insight into the latent switching mechanism, we also carry out the fluence-dependent ultrafast optical pump-probe reflectivity experiment to investigate coherent phonon oscillations. The data under different pump fluences are shown in Fig.\ref{Fig:5}(a), all the decay processes can be described by a double-exponential function to fit the decay process: $\Delta R/R=A_1 \exp \left(-t / \tau_{1}\right)+A_2 \exp \left(-t / \tau_{2}\right)+C$, Where $A_1$, $A_2$ in the formula represents the amplitude of the photo-induced reflectivity change, $\tau_{1}$, $\tau_{2}$ stands for the two fast relaxation time and $C$ is constant for long lived thermal diffusion process. After subtracting the background decay processes, only one A$_1$ phonon mode could be extracted from time domain coherent oscillation by Fourier transform. This A$_1$ mode phonon frequency is about 5.4 THz ($\sim$180 cm$^{-1}$) in Fig.\ref{Fig:5}(b) under low 130 $\mu$J/cm$^2$ pump excitation, which is in accord with the previous static Raman scattering measurements\cite{Wulferding2020}. The dependence of the phonon frequency on the pump fluence is depicted in Fig.\ref{Fig:5}(c). With the pump fluence further increasing, the A$_1$ phonon frequency in Fig.\ref{Fig:5}(c) shows distinct continuous red-shifted from 5.4 THz to 5.1 THz than equilibrium state ($\sim$5$\%$ frequency red-shifted), indicating a phonon mode softened behavior. Similar phonon softening behavior has been observed in previous fluence-dependent transient optical reflectivity experiments of tellurium(Te), bismuth(Bi) and tin selenide(SnSe) crystals\cite{Hunsche1995,PhysRevLett.88.067401,PhysRevX.12.011029}. This behavior could be regarded as an extra evidence for the structural distortion\cite{Fritz2007}. 
  
  In general, valence electrons and ionic cores can be regarded as distinct components that couple through the electron-phonon interaction. Previous hydrodynamic transport and Raman spectroscopy results indicated that the electron-phonon interaction was strong in $\beta$-WP$_2$\cite{Coulter2018,Osterhoudt2021}, similar to tellurium (Te) and antimony (Sb). Electronic ultrafast bond softening mechanism is usually taken into consideration for  lattice instability in these materials. The lattice structure is very sensitive to the population distribution of electrons within the conduction bands. The A$_1$ coherent optical phonon mode at symmetric zone center could be driven by light-induced charge carriers and the vibrational excitation is generally believed to be displacive\cite{Zeiger1992}. The population of valence electrons are redistributed after photo-excitation so that makes the bond softening through electron-phonon coupling. 
  With high excitation fluence, the interatomic forces that bind solids and the atoms positions can be altered compare to equilibrium state. The observed phonon softening is consistent with behavior related to the carriers-induced ultrafast bond softening \cite{Fritz2007}. Perturbation by light makes structural instability in the xz plane connected to the polar axis, resulting SHG and dominant $\chi_{zxx}$ reduction as the previous high-pressure strain manipulation effect in TaAs SHG response and dominant $\chi_{zzz}$\cite{Li2022}. However, since time-dependent optical reflectivity monitors the A$_1$ vibration mode indirectly and can not provide messages about atomic positions, detailed femtosecond x-ray diffraction is required to elucidate the dynamics of the high-amplitude phonon and to proof this mechanism.  For topological semimetals, where structural symmetry is intimately tied to topology, other possibility is that the band structure transformation is also induced by light, resulting SHG response change. To determine the origin of the specific ultrafast symmetry switching, further time-resolved angle-resolved photoemission spectroscopy (tr-ARPES) is also indispensable. Since very strong strength of interlayer coupling in WP$_2$, it is certain that the essence of driving mechanism is entirely different with the shear displacements in layered T$_d$-XTe$_2$ (X= W,Mo).

\section{CONCLUSION}
In summary, we performed combined static and time-resolved SHG measurements in Type-\uppercase\expandafter{\romannumeral2} Weyl semimetal $\beta$-WP$_2$. As a topological polar metal with extremely high conductivity, WP$_2$ exhibits rather strong SHG responses that maximum $|\chi_{xxz}|$ is about 150 pV/m, which is one or two order larger than ferroelectric LiNbO$_3$, LiTaO$_3$ and polar metal LiOsO$_3$ with 1.55 eV photon excitation. The static SHG results enrich the visions of intriguing polar metal nonlinear optics and enable potentially practical applications for topological optoelectronic devices. After photo-excitation, intense laser pulse could optically drive novel symmetry switching and tune nonlinear optical responses on ultrafast timescales. The results indicate that $\beta$-WP$_2$ have strong tendency to the topological trivial state. The ultrafast switching is accompanied by a distinct softening of the lowest-frequency A$_1$ phonon. The essence of driving mechanism is entirely different with the T$_d$ phase WTe$_2$/MoTe$_2$ light-induced shear displacements. Our study provides explicit evidence of ultrafast symmetry switching, which opens up experimental possibilities for ultrafast manipulation of the topological nonlinear optics properties of solids. This ultrafast symmetry switching will have wide ranging applications, particularly for topological semimetals, where symmetry is intimately tied to topology. Further time-resolved x-ray diffraction and time-resolved angle-resolved photoelectron spectroscopy is pursued for the transient lattice and topological electronic structure switching in future.

\noindent$\large{\textbf{Methods}}$

\noindent Crystal growth and structure characterization.

Single crystals of $\beta$-WP$_2$ were grown via chemical vapor transport with iodine as the transport agent. All capsulation processes were operated in a sealed glove box. Detailed growth method could be found in reference\cite{Su2019}. The as-grown three-dimensional crystals with needle-shaped exhibit multiple shiny surface facets. $\beta$-WP$_2$ is air stable. Crystal structure and crystal orientation were confirmed using single crystal X-ray diffractometer (SXRD) Bruker D8 Venture with Mo-K$_\alpha$ radiation ($\lambda$= 0.71 $\AA$) at room temperature. The determined lattice parameters were consist with the previous report in the literature. 

~\\
\noindent Optics set-up for reflecting static/time-resolved second-harmonic generation and pump-probed reflectively measurement

The optics set-up for measuring reflecting static/time-resolved SHG is depicted in the Fig.\ref{Fig:1}b. The measurements were performed in a Ti:sapphire amplified laser (Spitfire Ace) system of 800-nm centre-wavelength pulses for a duration of 35 fs (1 kHz repetition rate). Generator pulses pass through a mechanical chopper that provides amplitude modulation at 377 Hz. And the pulses are focused at 11°/45° incidence onto the sample. The spot sizes of pump and probe are measured about 150,120 $\mu$m in 45° incidence configuration by a 100-$\mu$m diameter pinhole, respectively. The pump is in normal incidence relative to xz plane with a half waveplate to vary linear polarization. The probe fluence of the time-resolved SHG is set as 0.30 mJ/cm$^2$. A Quarter waveplate and a rotatable wired-polarizer in the incidence beam path are used to generate circular polarization and to vary the direction of linear polarization. Both the outgoing fundamental and the second-harmonic beams are directed to a short-pass, band-pass filter group that ensures only the second-harmonic light to reach the photomultiplier tube photodetector (PMM01, Thorlabs, Bias voltage set as 0.65V). Another rotatable wire-grid polarizer placed before the PMT allows for analysis of the polarization of the second-harmonic beam. Temperature-dependence measurements were performed by mounting the WP$_2$ sample on a holder of helium-free closed-cycle cryostat (Montana Instruments). Benchmark measurements on $\beta$-WP$_2$, ZnTe and GaAs were performed in ambient conditions and the samples mounted flat on a xyz-micrometer stage to maximize the signal at room temperature.

For fundamental time-dependent reflectivity measurement, the standard pump-probe scheme can be found elsewhere\cite{Hu2022}.

\begin{center}
\small{\textbf{ACKNOWLEDGMENTS}}
\end{center}

This work was supported by the National Natural Science Foundation of China (No. 11888101), the National Key Research and Development Program of China (No. 2017YFA0302904).

\bibliography{WP2}

\begin{thebibliography}{58}%
\makeatletter
\providecommand \@ifxundefined [1]{%
 \@ifx{#1\undefined}
}%
\providecommand \@ifnum [1]{%
 \ifnum #1\expandafter \@firstoftwo
 \else \expandafter \@secondoftwo
 \fi
}%
\providecommand \@ifx [1]{%
 \ifx #1\expandafter \@firstoftwo
 \else \expandafter \@secondoftwo
 \fi
}%
\providecommand \natexlab [1]{#1}%
\providecommand \enquote  [1]{``#1''}%
\providecommand \bibnamefont  [1]{#1}%
\providecommand \bibfnamefont [1]{#1}%
\providecommand \citenamefont [1]{#1}%
\providecommand \href@noop [0]{\@secondoftwo}%
\providecommand \href [0]{\begingroup \@sanitize@url \@href}%
\providecommand \@href[1]{\@@startlink{#1}\@@href}%
\providecommand \@@href[1]{\endgroup#1\@@endlink}%
\providecommand \@sanitize@url [0]{\catcode `\\12\catcode `\$12\catcode
  `\&12\catcode `\#12\catcode `\^12\catcode `\_12\catcode `\%12\relax}%
\providecommand \@@startlink[1]{}%
\providecommand \@@endlink[0]{}%
\providecommand \url  [0]{\begingroup\@sanitize@url \@url }%
\providecommand \@url [1]{\endgroup\@href {#1}{\urlprefix }}%
\providecommand \urlprefix  [0]{URL }%
\providecommand \Eprint [0]{\href }%
\providecommand \doibase [0]{http://dx.doi.org/}%
\providecommand \selectlanguage [0]{\@gobble}%
\providecommand \bibinfo  [0]{\@secondoftwo}%
\providecommand \bibfield  [0]{\@secondoftwo}%
\providecommand \translation [1]{[#1]}%
\providecommand \BibitemOpen [0]{}%
\providecommand \bibitemStop [0]{}%
\providecommand \bibitemNoStop [0]{.\EOS\space}%
\providecommand \EOS [0]{\spacefactor3000\relax}%
\providecommand \BibitemShut  [1]{\csname bibitem#1\endcsname}%
\let\auto@bib@innerbib\@empty
\bibitem [{\citenamefont {Weyl}(1929)}]{Weyl1929}%
  \BibitemOpen
  \bibfield  {author} {\bibinfo {author} {\bibfnamefont {H.}~\bibnamefont
  {Weyl}},\ }\href {\doibase 10.1007/BF01339504} {\bibfield  {journal}
  {\bibinfo  {journal} {Zeitschrift für Physik}\ }\textbf {\bibinfo {volume}
  {56}},\ \bibinfo {pages} {330} (\bibinfo {year} {1929})}\BibitemShut
  {NoStop}%
\bibitem [{\citenamefont {Xu}\ \emph {et~al.}(2016{\natexlab{a}})\citenamefont
  {Xu}, \citenamefont {Weng}, \citenamefont {Lv}, \citenamefont {Matt},
  \citenamefont {Park}, \citenamefont {Bisti}, \citenamefont {Strocov},
  \citenamefont {Gawryluk}, \citenamefont {Pomjakushina}, \citenamefont
  {Conder}, \citenamefont {Plumb}, \citenamefont {Radovic}, \citenamefont
  {Autès}, \citenamefont {Yazyev}, \citenamefont {Fang}, \citenamefont {Dai},
  \citenamefont {Qian}, \citenamefont {Mesot}, \citenamefont {Ding},\ and\
  \citenamefont {Shi}}]{Xu2016}%
  \BibitemOpen
  \bibfield  {author} {\bibinfo {author} {\bibfnamefont {N.}~\bibnamefont
  {Xu}}, \bibinfo {author} {\bibfnamefont {H.~M.}\ \bibnamefont {Weng}},
  \bibinfo {author} {\bibfnamefont {B.~Q.}\ \bibnamefont {Lv}}, \bibinfo
  {author} {\bibfnamefont {C.~E.}\ \bibnamefont {Matt}}, \bibinfo {author}
  {\bibfnamefont {J.}~\bibnamefont {Park}}, \bibinfo {author} {\bibfnamefont
  {F.}~\bibnamefont {Bisti}}, \bibinfo {author} {\bibfnamefont {V.~N.}\
  \bibnamefont {Strocov}}, \bibinfo {author} {\bibfnamefont {D.}~\bibnamefont
  {Gawryluk}}, \bibinfo {author} {\bibfnamefont {E.}~\bibnamefont
  {Pomjakushina}}, \bibinfo {author} {\bibfnamefont {K.}~\bibnamefont
  {Conder}}, \bibinfo {author} {\bibfnamefont {N.~C.}\ \bibnamefont {Plumb}},
  \bibinfo {author} {\bibfnamefont {M.}~\bibnamefont {Radovic}}, \bibinfo
  {author} {\bibfnamefont {G.}~\bibnamefont {Autès}}, \bibinfo {author}
  {\bibfnamefont {O.~V.}\ \bibnamefont {Yazyev}}, \bibinfo {author}
  {\bibfnamefont {Z.}~\bibnamefont {Fang}}, \bibinfo {author} {\bibfnamefont
  {X.}~\bibnamefont {Dai}}, \bibinfo {author} {\bibfnamefont {T.}~\bibnamefont
  {Qian}}, \bibinfo {author} {\bibfnamefont {J.}~\bibnamefont {Mesot}},
  \bibinfo {author} {\bibfnamefont {H.}~\bibnamefont {Ding}}, \ and\ \bibinfo
  {author} {\bibfnamefont {M.}~\bibnamefont {Shi}},\ }\href {\doibase
  10.1038/ncomms11006} {\bibfield  {journal} {\bibinfo  {journal} {Nature
  Communications}\ }\textbf {\bibinfo {volume} {7}},\ \bibinfo {pages} {11006}
  (\bibinfo {year} {2016}{\natexlab{a}})}\BibitemShut {NoStop}%
\bibitem [{\citenamefont {Yao}\ \emph {et~al.}(2019)\citenamefont {Yao},
  \citenamefont {Xu}, \citenamefont {Wu}, \citenamefont {Aut\`es},
  \citenamefont {Kumar}, \citenamefont {Strocov}, \citenamefont {Plumb},
  \citenamefont {Radovic}, \citenamefont {Yazyev}, \citenamefont {Felser},
  \citenamefont {Mesot},\ and\ \citenamefont {Shi}}]{Yao2019}%
  \BibitemOpen
  \bibfield  {author} {\bibinfo {author} {\bibfnamefont {M.-Y.}\ \bibnamefont
  {Yao}}, \bibinfo {author} {\bibfnamefont {N.}~\bibnamefont {Xu}}, \bibinfo
  {author} {\bibfnamefont {Q.~S.}\ \bibnamefont {Wu}}, \bibinfo {author}
  {\bibfnamefont {G.}~\bibnamefont {Aut\`es}}, \bibinfo {author} {\bibfnamefont
  {N.}~\bibnamefont {Kumar}}, \bibinfo {author} {\bibfnamefont {V.~N.}\
  \bibnamefont {Strocov}}, \bibinfo {author} {\bibfnamefont {N.~C.}\
  \bibnamefont {Plumb}}, \bibinfo {author} {\bibfnamefont {M.}~\bibnamefont
  {Radovic}}, \bibinfo {author} {\bibfnamefont {O.~V.}\ \bibnamefont {Yazyev}},
  \bibinfo {author} {\bibfnamefont {C.}~\bibnamefont {Felser}}, \bibinfo
  {author} {\bibfnamefont {J.}~\bibnamefont {Mesot}}, \ and\ \bibinfo {author}
  {\bibfnamefont {M.}~\bibnamefont {Shi}},\ }\href {\doibase
  10.1103/PhysRevLett.122.176402} {\bibfield  {journal} {\bibinfo  {journal}
  {Phys. Rev. Lett.}\ }\textbf {\bibinfo {volume} {122}},\ \bibinfo {pages}
  {176402} (\bibinfo {year} {2019})}\BibitemShut {NoStop}%
\bibitem [{\citenamefont {Xu}\ \emph {et~al.}(2015)\citenamefont {Xu},
  \citenamefont {Belopolski}, \citenamefont {Alidoust}, \citenamefont
  {Neupane}, \citenamefont {Bian}, \citenamefont {Zhang}, \citenamefont
  {Sankar}, \citenamefont {Chang}, \citenamefont {Yuan}, \citenamefont {Lee},
  \citenamefont {Huang}, \citenamefont {Zheng}, \citenamefont {Ma},
  \citenamefont {Sanchez}, \citenamefont {Wang}, \citenamefont {Bansil},
  \citenamefont {Chou}, \citenamefont {Shibayev}, \citenamefont {Lin},
  \citenamefont {Jia},\ and\ \citenamefont {Hasan}}]{Xu2015}%
  \BibitemOpen
  \bibfield  {author} {\bibinfo {author} {\bibfnamefont {S.-Y.}\ \bibnamefont
  {Xu}}, \bibinfo {author} {\bibfnamefont {I.}~\bibnamefont {Belopolski}},
  \bibinfo {author} {\bibfnamefont {N.}~\bibnamefont {Alidoust}}, \bibinfo
  {author} {\bibfnamefont {M.}~\bibnamefont {Neupane}}, \bibinfo {author}
  {\bibfnamefont {G.}~\bibnamefont {Bian}}, \bibinfo {author} {\bibfnamefont
  {C.}~\bibnamefont {Zhang}}, \bibinfo {author} {\bibfnamefont
  {R.}~\bibnamefont {Sankar}}, \bibinfo {author} {\bibfnamefont
  {G.}~\bibnamefont {Chang}}, \bibinfo {author} {\bibfnamefont
  {Z.}~\bibnamefont {Yuan}}, \bibinfo {author} {\bibfnamefont {C.-C.}\
  \bibnamefont {Lee}}, \bibinfo {author} {\bibfnamefont {S.-M.}\ \bibnamefont
  {Huang}}, \bibinfo {author} {\bibfnamefont {H.}~\bibnamefont {Zheng}},
  \bibinfo {author} {\bibfnamefont {J.}~\bibnamefont {Ma}}, \bibinfo {author}
  {\bibfnamefont {D.~S.}\ \bibnamefont {Sanchez}}, \bibinfo {author}
  {\bibfnamefont {B.}~\bibnamefont {Wang}}, \bibinfo {author} {\bibfnamefont
  {A.}~\bibnamefont {Bansil}}, \bibinfo {author} {\bibfnamefont
  {F.}~\bibnamefont {Chou}}, \bibinfo {author} {\bibfnamefont {P.~P.}\
  \bibnamefont {Shibayev}}, \bibinfo {author} {\bibfnamefont {H.}~\bibnamefont
  {Lin}}, \bibinfo {author} {\bibfnamefont {S.}~\bibnamefont {Jia}}, \ and\
  \bibinfo {author} {\bibfnamefont {M.~Z.}\ \bibnamefont {Hasan}},\ }\href
  {\doibase 10.1126/science.aaa9297} {\bibfield  {journal} {\bibinfo  {journal}
  {Science}\ }\textbf {\bibinfo {volume} {349}},\ \bibinfo {pages} {613}
  (\bibinfo {year} {2015})},\ \Eprint
  {http://arxiv.org/abs/https://www.science.org/doi/pdf/10.1126/science.aaa9297}
  {https://www.science.org/doi/pdf/10.1126/science.aaa9297} \BibitemShut
  {NoStop}%
\bibitem [{\citenamefont {Lv}\ \emph {et~al.}(2015)\citenamefont {Lv},
  \citenamefont {Weng}, \citenamefont {Fu}, \citenamefont {Wang}, \citenamefont
  {Miao}, \citenamefont {Ma}, \citenamefont {Richard}, \citenamefont {Huang},
  \citenamefont {Zhao}, \citenamefont {Chen}, \citenamefont {Fang},
  \citenamefont {Dai}, \citenamefont {Qian},\ and\ \citenamefont
  {Ding}}]{Lv2015}%
  \BibitemOpen
  \bibfield  {author} {\bibinfo {author} {\bibfnamefont {B.~Q.}\ \bibnamefont
  {Lv}}, \bibinfo {author} {\bibfnamefont {H.~M.}\ \bibnamefont {Weng}},
  \bibinfo {author} {\bibfnamefont {B.~B.}\ \bibnamefont {Fu}}, \bibinfo
  {author} {\bibfnamefont {X.~P.}\ \bibnamefont {Wang}}, \bibinfo {author}
  {\bibfnamefont {H.}~\bibnamefont {Miao}}, \bibinfo {author} {\bibfnamefont
  {J.}~\bibnamefont {Ma}}, \bibinfo {author} {\bibfnamefont {P.}~\bibnamefont
  {Richard}}, \bibinfo {author} {\bibfnamefont {X.~C.}\ \bibnamefont {Huang}},
  \bibinfo {author} {\bibfnamefont {L.~X.}\ \bibnamefont {Zhao}}, \bibinfo
  {author} {\bibfnamefont {G.~F.}\ \bibnamefont {Chen}}, \bibinfo {author}
  {\bibfnamefont {Z.}~\bibnamefont {Fang}}, \bibinfo {author} {\bibfnamefont
  {X.}~\bibnamefont {Dai}}, \bibinfo {author} {\bibfnamefont {T.}~\bibnamefont
  {Qian}}, \ and\ \bibinfo {author} {\bibfnamefont {H.}~\bibnamefont {Ding}},\
  }\href {\doibase 10.1103/PhysRevX.5.031013} {\bibfield  {journal} {\bibinfo
  {journal} {Phys. Rev. X}\ }\textbf {\bibinfo {volume} {5}},\ \bibinfo {pages}
  {031013} (\bibinfo {year} {2015})}\BibitemShut {NoStop}%
\bibitem [{\citenamefont {Weng}\ \emph {et~al.}(2015)\citenamefont {Weng},
  \citenamefont {Fang}, \citenamefont {Fang}, \citenamefont {Bernevig},\ and\
  \citenamefont {Dai}}]{PhysRevX.5.011029}%
  \BibitemOpen
  \bibfield  {author} {\bibinfo {author} {\bibfnamefont {H.}~\bibnamefont
  {Weng}}, \bibinfo {author} {\bibfnamefont {C.}~\bibnamefont {Fang}}, \bibinfo
  {author} {\bibfnamefont {Z.}~\bibnamefont {Fang}}, \bibinfo {author}
  {\bibfnamefont {B.~A.}\ \bibnamefont {Bernevig}}, \ and\ \bibinfo {author}
  {\bibfnamefont {X.}~\bibnamefont {Dai}},\ }\href {\doibase
  10.1103/PhysRevX.5.011029} {\bibfield  {journal} {\bibinfo  {journal} {Phys.
  Rev. X}\ }\textbf {\bibinfo {volume} {5}},\ \bibinfo {pages} {011029}
  (\bibinfo {year} {2015})}\BibitemShut {NoStop}%
\bibitem [{\citenamefont {Xu}\ \emph {et~al.}(2020)\citenamefont {Xu},
  \citenamefont {Zhao}, \citenamefont {Yi}, \citenamefont {Wang}, \citenamefont
  {Yin}, \citenamefont {Wang}, \citenamefont {Hu}, \citenamefont {Wang},
  \citenamefont {Liu}, \citenamefont {Xu}, \citenamefont {Lu}, \citenamefont
  {Soluyanov}, \citenamefont {Lei}, \citenamefont {Shi}, \citenamefont {Luo},\
  and\ \citenamefont {Chen}}]{Xu2020}%
  \BibitemOpen
  \bibfield  {author} {\bibinfo {author} {\bibfnamefont {Y.}~\bibnamefont
  {Xu}}, \bibinfo {author} {\bibfnamefont {J.}~\bibnamefont {Zhao}}, \bibinfo
  {author} {\bibfnamefont {C.}~\bibnamefont {Yi}}, \bibinfo {author}
  {\bibfnamefont {Q.}~\bibnamefont {Wang}}, \bibinfo {author} {\bibfnamefont
  {Q.}~\bibnamefont {Yin}}, \bibinfo {author} {\bibfnamefont {Y.}~\bibnamefont
  {Wang}}, \bibinfo {author} {\bibfnamefont {X.}~\bibnamefont {Hu}}, \bibinfo
  {author} {\bibfnamefont {L.}~\bibnamefont {Wang}}, \bibinfo {author}
  {\bibfnamefont {E.}~\bibnamefont {Liu}}, \bibinfo {author} {\bibfnamefont
  {G.}~\bibnamefont {Xu}}, \bibinfo {author} {\bibfnamefont {L.}~\bibnamefont
  {Lu}}, \bibinfo {author} {\bibfnamefont {A.~A.}\ \bibnamefont {Soluyanov}},
  \bibinfo {author} {\bibfnamefont {H.}~\bibnamefont {Lei}}, \bibinfo {author}
  {\bibfnamefont {Y.}~\bibnamefont {Shi}}, \bibinfo {author} {\bibfnamefont
  {J.}~\bibnamefont {Luo}}, \ and\ \bibinfo {author} {\bibfnamefont {Z.-G.}\
  \bibnamefont {Chen}},\ }\href {\doibase 10.1038/s41467-020-17234-0}
  {\bibfield  {journal} {\bibinfo  {journal} {Nature Communications}\ }\textbf
  {\bibinfo {volume} {11}},\ \bibinfo {pages} {3985} (\bibinfo {year}
  {2020})}\BibitemShut {NoStop}%
\bibitem [{\citenamefont {Soluyanov}\ \emph {et~al.}(2015)\citenamefont
  {Soluyanov}, \citenamefont {Gresch}, \citenamefont {Wang}, \citenamefont
  {Wu}, \citenamefont {Troyer}, \citenamefont {Dai},\ and\ \citenamefont
  {Bernevig}}]{Soluyanov2015}%
  \BibitemOpen
  \bibfield  {author} {\bibinfo {author} {\bibfnamefont {A.~A.}\ \bibnamefont
  {Soluyanov}}, \bibinfo {author} {\bibfnamefont {D.}~\bibnamefont {Gresch}},
  \bibinfo {author} {\bibfnamefont {Z.}~\bibnamefont {Wang}}, \bibinfo {author}
  {\bibfnamefont {Q.}~\bibnamefont {Wu}}, \bibinfo {author} {\bibfnamefont
  {M.}~\bibnamefont {Troyer}}, \bibinfo {author} {\bibfnamefont
  {X.}~\bibnamefont {Dai}}, \ and\ \bibinfo {author} {\bibfnamefont {B.~A.}\
  \bibnamefont {Bernevig}},\ }\href {\doibase 10.1038/nature15768} {\bibfield
  {journal} {\bibinfo  {journal} {Nature}\ }\textbf {\bibinfo {volume} {527}},\
  \bibinfo {pages} {495} (\bibinfo {year} {2015})}\BibitemShut {NoStop}%
\bibitem [{\citenamefont {Armitage}\ \emph {et~al.}(2018)\citenamefont
  {Armitage}, \citenamefont {Mele},\ and\ \citenamefont
  {Vishwanath}}]{Armitage2018}%
  \BibitemOpen
  \bibfield  {author} {\bibinfo {author} {\bibfnamefont {N.~P.}\ \bibnamefont
  {Armitage}}, \bibinfo {author} {\bibfnamefont {E.~J.}\ \bibnamefont {Mele}},
  \ and\ \bibinfo {author} {\bibfnamefont {A.}~\bibnamefont {Vishwanath}},\
  }\href {\doibase 10.1103/RevModPhys.90.015001} {\bibfield  {journal}
  {\bibinfo  {journal} {Rev. Mod. Phys.}\ }\textbf {\bibinfo {volume} {90}},\
  \bibinfo {pages} {015001} (\bibinfo {year} {2018})}\BibitemShut {NoStop}%
\bibitem [{\citenamefont {Ali}\ \emph {et~al.}(2014)\citenamefont {Ali},
  \citenamefont {Xiong}, \citenamefont {Flynn}, \citenamefont {Tao},
  \citenamefont {Gibson}, \citenamefont {Schoop}, \citenamefont {Liang},
  \citenamefont {Haldolaarachchige}, \citenamefont {Hirschberger},
  \citenamefont {Ong},\ and\ \citenamefont {Cava}}]{Ali2014}%
  \BibitemOpen
  \bibfield  {author} {\bibinfo {author} {\bibfnamefont {M.~N.}\ \bibnamefont
  {Ali}}, \bibinfo {author} {\bibfnamefont {J.}~\bibnamefont {Xiong}}, \bibinfo
  {author} {\bibfnamefont {S.}~\bibnamefont {Flynn}}, \bibinfo {author}
  {\bibfnamefont {J.}~\bibnamefont {Tao}}, \bibinfo {author} {\bibfnamefont
  {Q.~D.}\ \bibnamefont {Gibson}}, \bibinfo {author} {\bibfnamefont {L.~M.}\
  \bibnamefont {Schoop}}, \bibinfo {author} {\bibfnamefont {T.}~\bibnamefont
  {Liang}}, \bibinfo {author} {\bibfnamefont {N.}~\bibnamefont
  {Haldolaarachchige}}, \bibinfo {author} {\bibfnamefont {M.}~\bibnamefont
  {Hirschberger}}, \bibinfo {author} {\bibfnamefont {N.~P.}\ \bibnamefont
  {Ong}}, \ and\ \bibinfo {author} {\bibfnamefont {R.~J.}\ \bibnamefont
  {Cava}},\ }\href {\doibase 10.1038/nature13763} {\bibfield  {journal}
  {\bibinfo  {journal} {Nature}\ }\textbf {\bibinfo {volume} {514}},\ \bibinfo
  {pages} {205} (\bibinfo {year} {2014})}\BibitemShut {NoStop}%
\bibitem [{\citenamefont {Kumar}\ \emph {et~al.}(2017)\citenamefont {Kumar},
  \citenamefont {Sun}, \citenamefont {Xu}, \citenamefont {Manna}, \citenamefont
  {Yao}, \citenamefont {Süss}, \citenamefont {Leermakers}, \citenamefont
  {Young}, \citenamefont {Förster}, \citenamefont {Schmidt}, \citenamefont
  {Borrmann}, \citenamefont {Yan}, \citenamefont {Zeitler}, \citenamefont
  {Shi}, \citenamefont {Felser},\ and\ \citenamefont {Shekhar}}]{Kumar2017}%
  \BibitemOpen
  \bibfield  {author} {\bibinfo {author} {\bibfnamefont {N.}~\bibnamefont
  {Kumar}}, \bibinfo {author} {\bibfnamefont {Y.}~\bibnamefont {Sun}}, \bibinfo
  {author} {\bibfnamefont {N.}~\bibnamefont {Xu}}, \bibinfo {author}
  {\bibfnamefont {K.}~\bibnamefont {Manna}}, \bibinfo {author} {\bibfnamefont
  {M.}~\bibnamefont {Yao}}, \bibinfo {author} {\bibfnamefont {V.}~\bibnamefont
  {Süss}}, \bibinfo {author} {\bibfnamefont {I.}~\bibnamefont {Leermakers}},
  \bibinfo {author} {\bibfnamefont {O.}~\bibnamefont {Young}}, \bibinfo
  {author} {\bibfnamefont {T.}~\bibnamefont {Förster}}, \bibinfo {author}
  {\bibfnamefont {M.}~\bibnamefont {Schmidt}}, \bibinfo {author} {\bibfnamefont
  {H.}~\bibnamefont {Borrmann}}, \bibinfo {author} {\bibfnamefont
  {B.}~\bibnamefont {Yan}}, \bibinfo {author} {\bibfnamefont {U.}~\bibnamefont
  {Zeitler}}, \bibinfo {author} {\bibfnamefont {M.}~\bibnamefont {Shi}},
  \bibinfo {author} {\bibfnamefont {C.}~\bibnamefont {Felser}}, \ and\ \bibinfo
  {author} {\bibfnamefont {C.}~\bibnamefont {Shekhar}},\ }\href {\doibase
  10.1038/s41467-017-01758-z} {\bibfield  {journal} {\bibinfo  {journal}
  {Nature Communications}\ }\textbf {\bibinfo {volume} {8}},\ \bibinfo {pages}
  {1642} (\bibinfo {year} {2017})}\BibitemShut {NoStop}%
\bibitem [{\citenamefont {Ma}\ \emph {et~al.}(2019)\citenamefont {Ma},
  \citenamefont {Gu}, \citenamefont {Liu}, \citenamefont {Lai}, \citenamefont
  {Yu}, \citenamefont {Zhuo}, \citenamefont {Liu}, \citenamefont {Chen},
  \citenamefont {Feng},\ and\ \citenamefont {Sun}}]{Ma2019}%
  \BibitemOpen
  \bibfield  {author} {\bibinfo {author} {\bibfnamefont {J.}~\bibnamefont
  {Ma}}, \bibinfo {author} {\bibfnamefont {Q.}~\bibnamefont {Gu}}, \bibinfo
  {author} {\bibfnamefont {Y.}~\bibnamefont {Liu}}, \bibinfo {author}
  {\bibfnamefont {J.}~\bibnamefont {Lai}}, \bibinfo {author} {\bibfnamefont
  {P.}~\bibnamefont {Yu}}, \bibinfo {author} {\bibfnamefont {X.}~\bibnamefont
  {Zhuo}}, \bibinfo {author} {\bibfnamefont {Z.}~\bibnamefont {Liu}}, \bibinfo
  {author} {\bibfnamefont {J.-H.}\ \bibnamefont {Chen}}, \bibinfo {author}
  {\bibfnamefont {J.}~\bibnamefont {Feng}}, \ and\ \bibinfo {author}
  {\bibfnamefont {D.}~\bibnamefont {Sun}},\ }\href {\doibase
  10.1038/s41563-019-0296-5} {\bibfield  {journal} {\bibinfo  {journal} {Nature
  Materials}\ }\textbf {\bibinfo {volume} {18}},\ \bibinfo {pages} {476}
  (\bibinfo {year} {2019})}\BibitemShut {NoStop}%
\bibitem [{\citenamefont {Chi}\ \emph {et~al.}(2018)\citenamefont {Chi},
  \citenamefont {Li}, \citenamefont {Xie}, \citenamefont {Zhao}, \citenamefont
  {Wang}, \citenamefont {Li}, \citenamefont {Yu}, \citenamefont {Wang},
  \citenamefont {Weng}, \citenamefont {Zhang},\ and\ \citenamefont
  {Wang}}]{Chi2018}%
  \BibitemOpen
  \bibfield  {author} {\bibinfo {author} {\bibfnamefont {S.}~\bibnamefont
  {Chi}}, \bibinfo {author} {\bibfnamefont {Z.}~\bibnamefont {Li}}, \bibinfo
  {author} {\bibfnamefont {Y.}~\bibnamefont {Xie}}, \bibinfo {author}
  {\bibfnamefont {Y.}~\bibnamefont {Zhao}}, \bibinfo {author} {\bibfnamefont
  {Z.}~\bibnamefont {Wang}}, \bibinfo {author} {\bibfnamefont {L.}~\bibnamefont
  {Li}}, \bibinfo {author} {\bibfnamefont {H.}~\bibnamefont {Yu}}, \bibinfo
  {author} {\bibfnamefont {G.}~\bibnamefont {Wang}}, \bibinfo {author}
  {\bibfnamefont {H.}~\bibnamefont {Weng}}, \bibinfo {author} {\bibfnamefont
  {H.}~\bibnamefont {Zhang}}, \ and\ \bibinfo {author} {\bibfnamefont
  {J.}~\bibnamefont {Wang}},\ }\href {\doibase
  https://doi.org/10.1002/adma.201801372} {\bibfield  {journal} {\bibinfo
  {journal} {Advanced Materials}\ }\textbf {\bibinfo {volume} {30}},\ \bibinfo
  {pages} {1801372} (\bibinfo {year} {2018})},\ \Eprint
  {http://arxiv.org/abs/https://onlinelibrary.wiley.com/doi/pdf/10.1002/adma.201801372}
  {https://onlinelibrary.wiley.com/doi/pdf/10.1002/adma.201801372} \BibitemShut
  {NoStop}%
\bibitem [{\citenamefont {Liu}\ \emph {et~al.}(2020)\citenamefont {Liu},
  \citenamefont {Xia}, \citenamefont {Xiao}, \citenamefont {García~de Abajo},\
  and\ \citenamefont {Sun}}]{Liu2020}%
  \BibitemOpen
  \bibfield  {author} {\bibinfo {author} {\bibfnamefont {J.}~\bibnamefont
  {Liu}}, \bibinfo {author} {\bibfnamefont {F.}~\bibnamefont {Xia}}, \bibinfo
  {author} {\bibfnamefont {D.}~\bibnamefont {Xiao}}, \bibinfo {author}
  {\bibfnamefont {F.~J.}\ \bibnamefont {García~de Abajo}}, \ and\ \bibinfo
  {author} {\bibfnamefont {D.}~\bibnamefont {Sun}},\ }\href {\doibase
  10.1038/s41563-020-0715-7} {\bibfield  {journal} {\bibinfo  {journal} {Nature
  Materials}\ }\textbf {\bibinfo {volume} {19}},\ \bibinfo {pages} {830}
  (\bibinfo {year} {2020})}\BibitemShut {NoStop}%
\bibitem [{\citenamefont {Zhang}\ \emph {et~al.}(2020)\citenamefont {Zhang},
  \citenamefont {Du}, \citenamefont {Qi}, \citenamefont {Cheng}, \citenamefont
  {Fan}, \citenamefont {Wei}, \citenamefont {Li}, \citenamefont {Wang},
  \citenamefont {Yu}, \citenamefont {Hu}, \citenamefont {Sun}, \citenamefont
  {Huang}, \citenamefont {Chu}, \citenamefont {Wan},\ and\ \citenamefont
  {Zeng}}]{Zhang2020}%
  \BibitemOpen
  \bibfield  {author} {\bibinfo {author} {\bibfnamefont {K.}~\bibnamefont
  {Zhang}}, \bibinfo {author} {\bibfnamefont {Y.}~\bibnamefont {Du}}, \bibinfo
  {author} {\bibfnamefont {Z.}~\bibnamefont {Qi}}, \bibinfo {author}
  {\bibfnamefont {B.}~\bibnamefont {Cheng}}, \bibinfo {author} {\bibfnamefont
  {X.}~\bibnamefont {Fan}}, \bibinfo {author} {\bibfnamefont {L.}~\bibnamefont
  {Wei}}, \bibinfo {author} {\bibfnamefont {L.}~\bibnamefont {Li}}, \bibinfo
  {author} {\bibfnamefont {D.}~\bibnamefont {Wang}}, \bibinfo {author}
  {\bibfnamefont {G.}~\bibnamefont {Yu}}, \bibinfo {author} {\bibfnamefont
  {S.}~\bibnamefont {Hu}}, \bibinfo {author} {\bibfnamefont {C.}~\bibnamefont
  {Sun}}, \bibinfo {author} {\bibfnamefont {Z.}~\bibnamefont {Huang}}, \bibinfo
  {author} {\bibfnamefont {J.}~\bibnamefont {Chu}}, \bibinfo {author}
  {\bibfnamefont {X.}~\bibnamefont {Wan}}, \ and\ \bibinfo {author}
  {\bibfnamefont {C.}~\bibnamefont {Zeng}},\ }\href {\doibase
  10.1103/PhysRevApplied.13.014058} {\bibfield  {journal} {\bibinfo  {journal}
  {Phys. Rev. Applied}\ }\textbf {\bibinfo {volume} {13}},\ \bibinfo {pages}
  {014058} (\bibinfo {year} {2020})}\BibitemShut {NoStop}%
\bibitem [{\citenamefont {Wu}\ \emph {et~al.}(2017)\citenamefont {Wu},
  \citenamefont {Patankar}, \citenamefont {Morimoto}, \citenamefont {Nair},
  \citenamefont {Thewalt}, \citenamefont {Little}, \citenamefont {Analytis},
  \citenamefont {Moore},\ and\ \citenamefont {Orenstein}}]{Wu2017}%
  \BibitemOpen
  \bibfield  {author} {\bibinfo {author} {\bibfnamefont {L.}~\bibnamefont
  {Wu}}, \bibinfo {author} {\bibfnamefont {S.}~\bibnamefont {Patankar}},
  \bibinfo {author} {\bibfnamefont {T.}~\bibnamefont {Morimoto}}, \bibinfo
  {author} {\bibfnamefont {N.~L.}\ \bibnamefont {Nair}}, \bibinfo {author}
  {\bibfnamefont {E.}~\bibnamefont {Thewalt}}, \bibinfo {author} {\bibfnamefont
  {A.}~\bibnamefont {Little}}, \bibinfo {author} {\bibfnamefont {J.~G.}\
  \bibnamefont {Analytis}}, \bibinfo {author} {\bibfnamefont {J.~E.}\
  \bibnamefont {Moore}}, \ and\ \bibinfo {author} {\bibfnamefont
  {J.}~\bibnamefont {Orenstein}},\ }\href {\doibase 10.1038/nphys3969}
  {\bibfield  {journal} {\bibinfo  {journal} {Nature Physics}\ }\textbf
  {\bibinfo {volume} {13}},\ \bibinfo {pages} {350} (\bibinfo {year}
  {2017})}\BibitemShut {NoStop}%
\bibitem [{\citenamefont {Osterhoudt}\ \emph {et~al.}(2019)\citenamefont
  {Osterhoudt}, \citenamefont {Diebel}, \citenamefont {Gray}, \citenamefont
  {Yang}, \citenamefont {Stanco}, \citenamefont {Huang}, \citenamefont {Shen},
  \citenamefont {Ni}, \citenamefont {Moll}, \citenamefont {Ran},\ and\
  \citenamefont {Burch}}]{Osterhoudt2019}%
  \BibitemOpen
  \bibfield  {author} {\bibinfo {author} {\bibfnamefont {G.~B.}\ \bibnamefont
  {Osterhoudt}}, \bibinfo {author} {\bibfnamefont {L.~K.}\ \bibnamefont
  {Diebel}}, \bibinfo {author} {\bibfnamefont {M.~J.}\ \bibnamefont {Gray}},
  \bibinfo {author} {\bibfnamefont {X.}~\bibnamefont {Yang}}, \bibinfo {author}
  {\bibfnamefont {J.}~\bibnamefont {Stanco}}, \bibinfo {author} {\bibfnamefont
  {X.}~\bibnamefont {Huang}}, \bibinfo {author} {\bibfnamefont
  {B.}~\bibnamefont {Shen}}, \bibinfo {author} {\bibfnamefont {N.}~\bibnamefont
  {Ni}}, \bibinfo {author} {\bibfnamefont {P.~J.~W.}\ \bibnamefont {Moll}},
  \bibinfo {author} {\bibfnamefont {Y.}~\bibnamefont {Ran}}, \ and\ \bibinfo
  {author} {\bibfnamefont {K.~S.}\ \bibnamefont {Burch}},\ }\href {\doibase
  10.1038/s41563-019-0297-4} {\bibfield  {journal} {\bibinfo  {journal} {Nature
  Materials}\ }\textbf {\bibinfo {volume} {18}},\ \bibinfo {pages} {471}
  (\bibinfo {year} {2019})}\BibitemShut {NoStop}%
\bibitem [{\citenamefont {Sirica}\ \emph {et~al.}(2019)\citenamefont {Sirica},
  \citenamefont {Tobey}, \citenamefont {Zhao}, \citenamefont {Chen},
  \citenamefont {Xu}, \citenamefont {Yang}, \citenamefont {Shen}, \citenamefont
  {Yarotski}, \citenamefont {Bowlan}, \citenamefont {Trugman}, \citenamefont
  {Zhu}, \citenamefont {Dai}, \citenamefont {Azad}, \citenamefont {Ni},
  \citenamefont {Qiu}, \citenamefont {Taylor},\ and\ \citenamefont
  {Prasankumar}}]{Sirica2019}%
  \BibitemOpen
  \bibfield  {author} {\bibinfo {author} {\bibfnamefont {N.}~\bibnamefont
  {Sirica}}, \bibinfo {author} {\bibfnamefont {R.~I.}\ \bibnamefont {Tobey}},
  \bibinfo {author} {\bibfnamefont {L.~X.}\ \bibnamefont {Zhao}}, \bibinfo
  {author} {\bibfnamefont {G.~F.}\ \bibnamefont {Chen}}, \bibinfo {author}
  {\bibfnamefont {B.}~\bibnamefont {Xu}}, \bibinfo {author} {\bibfnamefont
  {R.}~\bibnamefont {Yang}}, \bibinfo {author} {\bibfnamefont {B.}~\bibnamefont
  {Shen}}, \bibinfo {author} {\bibfnamefont {D.~A.}\ \bibnamefont {Yarotski}},
  \bibinfo {author} {\bibfnamefont {P.}~\bibnamefont {Bowlan}}, \bibinfo
  {author} {\bibfnamefont {S.~A.}\ \bibnamefont {Trugman}}, \bibinfo {author}
  {\bibfnamefont {J.-X.}\ \bibnamefont {Zhu}}, \bibinfo {author} {\bibfnamefont
  {Y.~M.}\ \bibnamefont {Dai}}, \bibinfo {author} {\bibfnamefont {A.~K.}\
  \bibnamefont {Azad}}, \bibinfo {author} {\bibfnamefont {N.}~\bibnamefont
  {Ni}}, \bibinfo {author} {\bibfnamefont {X.~G.}\ \bibnamefont {Qiu}},
  \bibinfo {author} {\bibfnamefont {A.~J.}\ \bibnamefont {Taylor}}, \ and\
  \bibinfo {author} {\bibfnamefont {R.~P.}\ \bibnamefont {Prasankumar}},\
  }\href {\doibase 10.1103/PhysRevLett.122.197401} {\bibfield  {journal}
  {\bibinfo  {journal} {Phys. Rev. Lett.}\ }\textbf {\bibinfo {volume} {122}},\
  \bibinfo {pages} {197401} (\bibinfo {year} {2019})}\BibitemShut {NoStop}%
\bibitem [{\citenamefont {Gao}\ \emph {et~al.}(2020)\citenamefont {Gao},
  \citenamefont {Kaushik}, \citenamefont {Philip}, \citenamefont {Li},
  \citenamefont {Qin}, \citenamefont {Liu}, \citenamefont {Zhang},
  \citenamefont {Su}, \citenamefont {Chen}, \citenamefont {Weng}, \citenamefont
  {Kharzeev}, \citenamefont {Liu},\ and\ \citenamefont {Qi}}]{Gao2020}%
  \BibitemOpen
  \bibfield  {author} {\bibinfo {author} {\bibfnamefont {Y.}~\bibnamefont
  {Gao}}, \bibinfo {author} {\bibfnamefont {S.}~\bibnamefont {Kaushik}},
  \bibinfo {author} {\bibfnamefont {E.~J.}\ \bibnamefont {Philip}}, \bibinfo
  {author} {\bibfnamefont {Z.}~\bibnamefont {Li}}, \bibinfo {author}
  {\bibfnamefont {Y.}~\bibnamefont {Qin}}, \bibinfo {author} {\bibfnamefont
  {Y.~P.}\ \bibnamefont {Liu}}, \bibinfo {author} {\bibfnamefont {W.~L.}\
  \bibnamefont {Zhang}}, \bibinfo {author} {\bibfnamefont {Y.~L.}\ \bibnamefont
  {Su}}, \bibinfo {author} {\bibfnamefont {X.}~\bibnamefont {Chen}}, \bibinfo
  {author} {\bibfnamefont {H.}~\bibnamefont {Weng}}, \bibinfo {author}
  {\bibfnamefont {D.~E.}\ \bibnamefont {Kharzeev}}, \bibinfo {author}
  {\bibfnamefont {M.~K.}\ \bibnamefont {Liu}}, \ and\ \bibinfo {author}
  {\bibfnamefont {J.}~\bibnamefont {Qi}},\ }\href {\doibase
  10.1038/s41467-020-14463-1} {\bibfield  {journal} {\bibinfo  {journal}
  {Nature Communications}\ }\textbf {\bibinfo {volume} {11}},\ \bibinfo {pages}
  {720} (\bibinfo {year} {2020})}\BibitemShut {NoStop}%
\bibitem [{\citenamefont {Sun}\ \emph {et~al.}(2015)\citenamefont {Sun},
  \citenamefont {Wu}, \citenamefont {Ali}, \citenamefont {Felser},\ and\
  \citenamefont {Yan}}]{Sun2015}%
  \BibitemOpen
  \bibfield  {author} {\bibinfo {author} {\bibfnamefont {Y.}~\bibnamefont
  {Sun}}, \bibinfo {author} {\bibfnamefont {S.-C.}\ \bibnamefont {Wu}},
  \bibinfo {author} {\bibfnamefont {M.~N.}\ \bibnamefont {Ali}}, \bibinfo
  {author} {\bibfnamefont {C.}~\bibnamefont {Felser}}, \ and\ \bibinfo {author}
  {\bibfnamefont {B.}~\bibnamefont {Yan}},\ }\href {\doibase
  10.1103/PhysRevB.92.161107} {\bibfield  {journal} {\bibinfo  {journal} {Phys.
  Rev. B}\ }\textbf {\bibinfo {volume} {92}},\ \bibinfo {pages} {161107}
  (\bibinfo {year} {2015})}\BibitemShut {NoStop}%
\bibitem [{\citenamefont {Aut\`es}\ \emph {et~al.}(2016)\citenamefont
  {Aut\`es}, \citenamefont {Gresch}, \citenamefont {Troyer}, \citenamefont
  {Soluyanov},\ and\ \citenamefont {Yazyev}}]{Autes2016}%
  \BibitemOpen
  \bibfield  {author} {\bibinfo {author} {\bibfnamefont {G.}~\bibnamefont
  {Aut\`es}}, \bibinfo {author} {\bibfnamefont {D.}~\bibnamefont {Gresch}},
  \bibinfo {author} {\bibfnamefont {M.}~\bibnamefont {Troyer}}, \bibinfo
  {author} {\bibfnamefont {A.~A.}\ \bibnamefont {Soluyanov}}, \ and\ \bibinfo
  {author} {\bibfnamefont {O.~V.}\ \bibnamefont {Yazyev}},\ }\href {\doibase
  10.1103/PhysRevLett.117.066402} {\bibfield  {journal} {\bibinfo  {journal}
  {Phys. Rev. Lett.}\ }\textbf {\bibinfo {volume} {117}},\ \bibinfo {pages}
  {066402} (\bibinfo {year} {2016})}\BibitemShut {NoStop}%
\bibitem [{\citenamefont {Drueke}\ \emph {et~al.}(2021)\citenamefont {Drueke},
  \citenamefont {Yang},\ and\ \citenamefont {Zhao}}]{Drueke2021}%
  \BibitemOpen
  \bibfield  {author} {\bibinfo {author} {\bibfnamefont {E.}~\bibnamefont
  {Drueke}}, \bibinfo {author} {\bibfnamefont {J.}~\bibnamefont {Yang}}, \ and\
  \bibinfo {author} {\bibfnamefont {L.}~\bibnamefont {Zhao}},\ }\href {\doibase
  10.1103/PhysRevB.104.064304} {\bibfield  {journal} {\bibinfo  {journal}
  {Phys. Rev. B}\ }\textbf {\bibinfo {volume} {104}},\ \bibinfo {pages}
  {064304} (\bibinfo {year} {2021})}\BibitemShut {NoStop}%
\bibitem [{\citenamefont {Lv}\ \emph {et~al.}(2021)\citenamefont {Lv},
  \citenamefont {Xu}, \citenamefont {Han}, \citenamefont {Zhang}, \citenamefont
  {Han}, \citenamefont {Zhou}, \citenamefont {Yao}, \citenamefont {Liu},
  \citenamefont {Lu}, \citenamefont {Weng}, \citenamefont {Xie}, \citenamefont
  {Chen}, \citenamefont {Hu}, \citenamefont {Chen},\ and\ \citenamefont
  {Zhu}}]{Lv2021}%
  \BibitemOpen
  \bibfield  {author} {\bibinfo {author} {\bibfnamefont {Y.-Y.}\ \bibnamefont
  {Lv}}, \bibinfo {author} {\bibfnamefont {J.}~\bibnamefont {Xu}}, \bibinfo
  {author} {\bibfnamefont {S.}~\bibnamefont {Han}}, \bibinfo {author}
  {\bibfnamefont {C.}~\bibnamefont {Zhang}}, \bibinfo {author} {\bibfnamefont
  {Y.}~\bibnamefont {Han}}, \bibinfo {author} {\bibfnamefont {J.}~\bibnamefont
  {Zhou}}, \bibinfo {author} {\bibfnamefont {S.-H.}\ \bibnamefont {Yao}},
  \bibinfo {author} {\bibfnamefont {X.-P.}\ \bibnamefont {Liu}}, \bibinfo
  {author} {\bibfnamefont {M.-H.}\ \bibnamefont {Lu}}, \bibinfo {author}
  {\bibfnamefont {H.}~\bibnamefont {Weng}}, \bibinfo {author} {\bibfnamefont
  {Z.}~\bibnamefont {Xie}}, \bibinfo {author} {\bibfnamefont {Y.~B.}\
  \bibnamefont {Chen}}, \bibinfo {author} {\bibfnamefont {J.}~\bibnamefont
  {Hu}}, \bibinfo {author} {\bibfnamefont {Y.-F.}\ \bibnamefont {Chen}}, \ and\
  \bibinfo {author} {\bibfnamefont {S.}~\bibnamefont {Zhu}},\ }\href {\doibase
  10.1038/s41467-021-26766-y} {\bibfield  {journal} {\bibinfo  {journal}
  {Nature Communications}\ }\textbf {\bibinfo {volume} {12}},\ \bibinfo {pages}
  {6437} (\bibinfo {year} {2021})}\BibitemShut {NoStop}%
\bibitem [{\citenamefont {Hsieh}\ \emph {et~al.}(2011)\citenamefont {Hsieh},
  \citenamefont {McIver}, \citenamefont {Torchinsky}, \citenamefont {Gardner},
  \citenamefont {Lee},\ and\ \citenamefont {Gedik}}]{Hsieh2011}%
  \BibitemOpen
  \bibfield  {author} {\bibinfo {author} {\bibfnamefont {D.}~\bibnamefont
  {Hsieh}}, \bibinfo {author} {\bibfnamefont {J.~W.}\ \bibnamefont {McIver}},
  \bibinfo {author} {\bibfnamefont {D.~H.}\ \bibnamefont {Torchinsky}},
  \bibinfo {author} {\bibfnamefont {D.~R.}\ \bibnamefont {Gardner}}, \bibinfo
  {author} {\bibfnamefont {Y.~S.}\ \bibnamefont {Lee}}, \ and\ \bibinfo
  {author} {\bibfnamefont {N.}~\bibnamefont {Gedik}},\ }\href {\doibase
  10.1103/PhysRevLett.106.057401} {\bibfield  {journal} {\bibinfo  {journal}
  {Phys. Rev. Lett.}\ }\textbf {\bibinfo {volume} {106}},\ \bibinfo {pages}
  {057401} (\bibinfo {year} {2011})}\BibitemShut {NoStop}%
\bibitem [{\citenamefont {Zhao}\ \emph {et~al.}(2016)\citenamefont {Zhao},
  \citenamefont {Torchinsky}, \citenamefont {Chu}, \citenamefont {Ivanov},
  \citenamefont {Lifshitz}, \citenamefont {Flint}, \citenamefont {Qi},
  \citenamefont {Cao},\ and\ \citenamefont {Hsieh}}]{Zhao2016}%
  \BibitemOpen
  \bibfield  {author} {\bibinfo {author} {\bibfnamefont {L.}~\bibnamefont
  {Zhao}}, \bibinfo {author} {\bibfnamefont {D.~H.}\ \bibnamefont
  {Torchinsky}}, \bibinfo {author} {\bibfnamefont {H.}~\bibnamefont {Chu}},
  \bibinfo {author} {\bibfnamefont {V.}~\bibnamefont {Ivanov}}, \bibinfo
  {author} {\bibfnamefont {R.}~\bibnamefont {Lifshitz}}, \bibinfo {author}
  {\bibfnamefont {R.}~\bibnamefont {Flint}}, \bibinfo {author} {\bibfnamefont
  {T.}~\bibnamefont {Qi}}, \bibinfo {author} {\bibfnamefont {G.}~\bibnamefont
  {Cao}}, \ and\ \bibinfo {author} {\bibfnamefont {D.}~\bibnamefont {Hsieh}},\
  }\href {\doibase 10.1038/nphys3517} {\bibfield  {journal} {\bibinfo
  {journal} {Nature Physics}\ }\textbf {\bibinfo {volume} {12}},\ \bibinfo
  {pages} {32} (\bibinfo {year} {2016})}\BibitemShut {NoStop}%
\bibitem [{\citenamefont {Harter}\ \emph {et~al.}(2017)\citenamefont {Harter},
  \citenamefont {Zhao}, \citenamefont {Yan}, \citenamefont {Mandrus},\ and\
  \citenamefont {Hsieh}}]{Harter2017}%
  \BibitemOpen
  \bibfield  {author} {\bibinfo {author} {\bibfnamefont {J.~W.}\ \bibnamefont
  {Harter}}, \bibinfo {author} {\bibfnamefont {Z.~Y.}\ \bibnamefont {Zhao}},
  \bibinfo {author} {\bibfnamefont {J.-Q.}\ \bibnamefont {Yan}}, \bibinfo
  {author} {\bibfnamefont {D.~G.}\ \bibnamefont {Mandrus}}, \ and\ \bibinfo
  {author} {\bibfnamefont {D.}~\bibnamefont {Hsieh}},\ }\href {\doibase
  10.1126/science.aad1188} {\bibfield  {journal} {\bibinfo  {journal}
  {Science}\ }\textbf {\bibinfo {volume} {356}},\ \bibinfo {pages} {295}
  (\bibinfo {year} {2017})},\ \Eprint
  {http://arxiv.org/abs/https://www.science.org/doi/pdf/10.1126/science.aad1188}
  {https://www.science.org/doi/pdf/10.1126/science.aad1188} \BibitemShut
  {NoStop}%
\bibitem [{\citenamefont {Fiebig}\ \emph {et~al.}(2005)\citenamefont {Fiebig},
  \citenamefont {Pavlov},\ and\ \citenamefont {Pisarev}}]{Fiebig2005}%
  \BibitemOpen
  \bibfield  {author} {\bibinfo {author} {\bibfnamefont {M.}~\bibnamefont
  {Fiebig}}, \bibinfo {author} {\bibfnamefont {V.~V.}\ \bibnamefont {Pavlov}},
  \ and\ \bibinfo {author} {\bibfnamefont {R.~V.}\ \bibnamefont {Pisarev}},\
  }\href {\doibase 10.1364/JOSAB.22.000096} {\bibfield  {journal} {\bibinfo
  {journal} {Journal of the Optical Society of America B: Optical Physics}\
  }\textbf {\bibinfo {volume} {22}},\ \bibinfo {pages} {96 – 118} (\bibinfo
  {year} {2005})}\BibitemShut {NoStop}%
\bibitem [{\citenamefont {Bergfeld}\ and\ \citenamefont
  {Daum}(2003)}]{Bergfeld2003}%
  \BibitemOpen
  \bibfield  {author} {\bibinfo {author} {\bibfnamefont {S.}~\bibnamefont
  {Bergfeld}}\ and\ \bibinfo {author} {\bibfnamefont {W.}~\bibnamefont
  {Daum}},\ }\href {\doibase 10.1103/PhysRevLett.90.036801} {\bibfield
  {journal} {\bibinfo  {journal} {Phys. Rev. Lett.}\ }\textbf {\bibinfo
  {volume} {90}},\ \bibinfo {pages} {036801} (\bibinfo {year}
  {2003})}\BibitemShut {NoStop}%
\bibitem [{\citenamefont {Wagner}\ \emph {et~al.}(1998)\citenamefont {Wagner},
  \citenamefont {K\"uhnelt}, \citenamefont {Langbein},\ and\ \citenamefont
  {Hvam}}]{PhysRevB.58.10494}%
  \BibitemOpen
  \bibfield  {author} {\bibinfo {author} {\bibfnamefont {H.~P.}\ \bibnamefont
  {Wagner}}, \bibinfo {author} {\bibfnamefont {M.}~\bibnamefont {K\"uhnelt}},
  \bibinfo {author} {\bibfnamefont {W.}~\bibnamefont {Langbein}}, \ and\
  \bibinfo {author} {\bibfnamefont {J.~M.}\ \bibnamefont {Hvam}},\ }\href
  {\doibase 10.1103/PhysRevB.58.10494} {\bibfield  {journal} {\bibinfo
  {journal} {Phys. Rev. B}\ }\textbf {\bibinfo {volume} {58}},\ \bibinfo
  {pages} {10494} (\bibinfo {year} {1998})}\BibitemShut {NoStop}%
\bibitem [{\citenamefont {Sie}\ \emph {et~al.}(2019)\citenamefont {Sie},
  \citenamefont {Nyby}, \citenamefont {Pemmaraju}, \citenamefont {Park},
  \citenamefont {Shen}, \citenamefont {Yang}, \citenamefont {Hoffmann},
  \citenamefont {Ofori-Okai}, \citenamefont {Li}, \citenamefont {Reid},
  \citenamefont {Weathersby}, \citenamefont {Mannebach}, \citenamefont
  {Finney}, \citenamefont {Rhodes}, \citenamefont {Chenet}, \citenamefont
  {Antony}, \citenamefont {Balicas}, \citenamefont {Hone}, \citenamefont
  {Devereaux}, \citenamefont {Heinz}, \citenamefont {Wang},\ and\ \citenamefont
  {Lindenberg}}]{Sie2019}%
  \BibitemOpen
  \bibfield  {author} {\bibinfo {author} {\bibfnamefont {E.~J.}\ \bibnamefont
  {Sie}}, \bibinfo {author} {\bibfnamefont {C.~M.}\ \bibnamefont {Nyby}},
  \bibinfo {author} {\bibfnamefont {C.~D.}\ \bibnamefont {Pemmaraju}}, \bibinfo
  {author} {\bibfnamefont {S.~J.}\ \bibnamefont {Park}}, \bibinfo {author}
  {\bibfnamefont {X.}~\bibnamefont {Shen}}, \bibinfo {author} {\bibfnamefont
  {J.}~\bibnamefont {Yang}}, \bibinfo {author} {\bibfnamefont {M.~C.}\
  \bibnamefont {Hoffmann}}, \bibinfo {author} {\bibfnamefont {B.~K.}\
  \bibnamefont {Ofori-Okai}}, \bibinfo {author} {\bibfnamefont
  {R.}~\bibnamefont {Li}}, \bibinfo {author} {\bibfnamefont {A.~H.}\
  \bibnamefont {Reid}}, \bibinfo {author} {\bibfnamefont {S.}~\bibnamefont
  {Weathersby}}, \bibinfo {author} {\bibfnamefont {E.}~\bibnamefont
  {Mannebach}}, \bibinfo {author} {\bibfnamefont {N.}~\bibnamefont {Finney}},
  \bibinfo {author} {\bibfnamefont {D.}~\bibnamefont {Rhodes}}, \bibinfo
  {author} {\bibfnamefont {D.}~\bibnamefont {Chenet}}, \bibinfo {author}
  {\bibfnamefont {A.}~\bibnamefont {Antony}}, \bibinfo {author} {\bibfnamefont
  {L.}~\bibnamefont {Balicas}}, \bibinfo {author} {\bibfnamefont
  {J.}~\bibnamefont {Hone}}, \bibinfo {author} {\bibfnamefont {T.~P.}\
  \bibnamefont {Devereaux}}, \bibinfo {author} {\bibfnamefont {T.~F.}\
  \bibnamefont {Heinz}}, \bibinfo {author} {\bibfnamefont {X.}~\bibnamefont
  {Wang}}, \ and\ \bibinfo {author} {\bibfnamefont {A.~M.}\ \bibnamefont
  {Lindenberg}},\ }\href {\doibase 10.1038/s41586-018-0809-4} {\bibfield
  {journal} {\bibinfo  {journal} {Nature}\ }\textbf {\bibinfo {volume} {565}},\
  \bibinfo {pages} {61} (\bibinfo {year} {2019})}\BibitemShut {NoStop}%
\bibitem [{\citenamefont {Zhang}\ \emph {et~al.}(2019)\citenamefont {Zhang},
  \citenamefont {Wang}, \citenamefont {Li}, \citenamefont {Shi}, \citenamefont
  {Wu}, \citenamefont {Lin}, \citenamefont {Zhang}, \citenamefont {Liu},
  \citenamefont {Liu}, \citenamefont {Wang}, \citenamefont {Dong},\ and\
  \citenamefont {Wang}}]{Zhang2019}%
  \BibitemOpen
  \bibfield  {author} {\bibinfo {author} {\bibfnamefont {M.~Y.}\ \bibnamefont
  {Zhang}}, \bibinfo {author} {\bibfnamefont {Z.~X.}\ \bibnamefont {Wang}},
  \bibinfo {author} {\bibfnamefont {Y.~N.}\ \bibnamefont {Li}}, \bibinfo
  {author} {\bibfnamefont {L.~Y.}\ \bibnamefont {Shi}}, \bibinfo {author}
  {\bibfnamefont {D.}~\bibnamefont {Wu}}, \bibinfo {author} {\bibfnamefont
  {T.}~\bibnamefont {Lin}}, \bibinfo {author} {\bibfnamefont {S.~J.}\
  \bibnamefont {Zhang}}, \bibinfo {author} {\bibfnamefont {Y.~Q.}\ \bibnamefont
  {Liu}}, \bibinfo {author} {\bibfnamefont {Q.~M.}\ \bibnamefont {Liu}},
  \bibinfo {author} {\bibfnamefont {J.}~\bibnamefont {Wang}}, \bibinfo {author}
  {\bibfnamefont {T.}~\bibnamefont {Dong}}, \ and\ \bibinfo {author}
  {\bibfnamefont {N.~L.}\ \bibnamefont {Wang}},\ }\href {\doibase
  10.1103/PhysRevX.9.021036} {\bibfield  {journal} {\bibinfo  {journal} {Phys.
  Rev. X}\ }\textbf {\bibinfo {volume} {9}},\ \bibinfo {pages} {021036}
  (\bibinfo {year} {2019})}\BibitemShut {NoStop}%
\bibitem [{\citenamefont {Su}\ \emph {et~al.}(2019)\citenamefont {Su},
  \citenamefont {Song}, \citenamefont {Hou}, \citenamefont {Chen},
  \citenamefont {Zhao}, \citenamefont {Ma}, \citenamefont {Yang}, \citenamefont
  {Guo}, \citenamefont {Luo},\ and\ \citenamefont {Chen}}]{Su2019}%
  \BibitemOpen
  \bibfield  {author} {\bibinfo {author} {\bibfnamefont {B.}~\bibnamefont
  {Su}}, \bibinfo {author} {\bibfnamefont {Y.}~\bibnamefont {Song}}, \bibinfo
  {author} {\bibfnamefont {Y.}~\bibnamefont {Hou}}, \bibinfo {author}
  {\bibfnamefont {X.}~\bibnamefont {Chen}}, \bibinfo {author} {\bibfnamefont
  {J.}~\bibnamefont {Zhao}}, \bibinfo {author} {\bibfnamefont {Y.}~\bibnamefont
  {Ma}}, \bibinfo {author} {\bibfnamefont {Y.}~\bibnamefont {Yang}}, \bibinfo
  {author} {\bibfnamefont {J.}~\bibnamefont {Guo}}, \bibinfo {author}
  {\bibfnamefont {J.}~\bibnamefont {Luo}}, \ and\ \bibinfo {author}
  {\bibfnamefont {Z.-G.}\ \bibnamefont {Chen}},\ }\href {\doibase
  https://doi.org/10.1002/adma.201903498} {\bibfield  {journal} {\bibinfo
  {journal} {Advanced Materials}\ }\textbf {\bibinfo {volume} {31}},\ \bibinfo
  {pages} {1903498} (\bibinfo {year} {2019})},\ \Eprint
  {http://arxiv.org/abs/https://onlinelibrary.wiley.com/doi/pdf/10.1002/adma.201903498}
  {https://onlinelibrary.wiley.com/doi/pdf/10.1002/adma.201903498} \BibitemShut
  {NoStop}%
\bibitem [{\citenamefont {Wulferding}\ \emph {et~al.}(2020)\citenamefont
  {Wulferding}, \citenamefont {Lemmens}, \citenamefont {B\"uscher},
  \citenamefont {Schmeltzer}, \citenamefont {Felser},\ and\ \citenamefont
  {Shekhar}}]{Wulferding2020}%
  \BibitemOpen
  \bibfield  {author} {\bibinfo {author} {\bibfnamefont {D.}~\bibnamefont
  {Wulferding}}, \bibinfo {author} {\bibfnamefont {P.}~\bibnamefont {Lemmens}},
  \bibinfo {author} {\bibfnamefont {F.}~\bibnamefont {B\"uscher}}, \bibinfo
  {author} {\bibfnamefont {D.}~\bibnamefont {Schmeltzer}}, \bibinfo {author}
  {\bibfnamefont {C.}~\bibnamefont {Felser}}, \ and\ \bibinfo {author}
  {\bibfnamefont {C.}~\bibnamefont {Shekhar}},\ }\href {\doibase
  10.1103/PhysRevB.102.075116} {\bibfield  {journal} {\bibinfo  {journal}
  {Phys. Rev. B}\ }\textbf {\bibinfo {volume} {102}},\ \bibinfo {pages}
  {075116} (\bibinfo {year} {2020})}\BibitemShut {NoStop}%
\bibitem [{\citenamefont {Osterhoudt}\ \emph {et~al.}(2021)\citenamefont
  {Osterhoudt}, \citenamefont {Wang}, \citenamefont {Garcia}, \citenamefont
  {Plisson}, \citenamefont {Gooth}, \citenamefont {Felser}, \citenamefont
  {Narang},\ and\ \citenamefont {Burch}}]{Osterhoudt2021}%
  \BibitemOpen
  \bibfield  {author} {\bibinfo {author} {\bibfnamefont {G.~B.}\ \bibnamefont
  {Osterhoudt}}, \bibinfo {author} {\bibfnamefont {Y.}~\bibnamefont {Wang}},
  \bibinfo {author} {\bibfnamefont {C.~A.~C.}\ \bibnamefont {Garcia}}, \bibinfo
  {author} {\bibfnamefont {V.~M.}\ \bibnamefont {Plisson}}, \bibinfo {author}
  {\bibfnamefont {J.}~\bibnamefont {Gooth}}, \bibinfo {author} {\bibfnamefont
  {C.}~\bibnamefont {Felser}}, \bibinfo {author} {\bibfnamefont
  {P.}~\bibnamefont {Narang}}, \ and\ \bibinfo {author} {\bibfnamefont {K.~S.}\
  \bibnamefont {Burch}},\ }\href {\doibase 10.1103/PhysRevX.11.011017}
  {\bibfield  {journal} {\bibinfo  {journal} {Phys. Rev. X}\ }\textbf {\bibinfo
  {volume} {11}},\ \bibinfo {pages} {011017} (\bibinfo {year}
  {2021})}\BibitemShut {NoStop}%
\bibitem [{\citenamefont {Wang}\ \emph {et~al.}(2017)\citenamefont {Wang},
  \citenamefont {Graf}, \citenamefont {Liu}, \citenamefont {Du}, \citenamefont
  {Zheng}, \citenamefont {Lei},\ and\ \citenamefont
  {Petrovic}}]{PhysRevB.96.121107}%
  \BibitemOpen
  \bibfield  {author} {\bibinfo {author} {\bibfnamefont {A.}~\bibnamefont
  {Wang}}, \bibinfo {author} {\bibfnamefont {D.}~\bibnamefont {Graf}}, \bibinfo
  {author} {\bibfnamefont {Y.}~\bibnamefont {Liu}}, \bibinfo {author}
  {\bibfnamefont {Q.}~\bibnamefont {Du}}, \bibinfo {author} {\bibfnamefont
  {J.}~\bibnamefont {Zheng}}, \bibinfo {author} {\bibfnamefont
  {H.}~\bibnamefont {Lei}}, \ and\ \bibinfo {author} {\bibfnamefont
  {C.}~\bibnamefont {Petrovic}},\ }\href {\doibase 10.1103/PhysRevB.96.121107}
  {\bibfield  {journal} {\bibinfo  {journal} {Phys. Rev. B}\ }\textbf {\bibinfo
  {volume} {96}},\ \bibinfo {pages} {121107} (\bibinfo {year}
  {2017})}\BibitemShut {NoStop}%
\bibitem [{\citenamefont {Bloembergen}\ and\ \citenamefont
  {Pershan}(1962)}]{Bloembergen1962}%
  \BibitemOpen
  \bibfield  {author} {\bibinfo {author} {\bibfnamefont {N.}~\bibnamefont
  {Bloembergen}}\ and\ \bibinfo {author} {\bibfnamefont {P.~S.}\ \bibnamefont
  {Pershan}},\ }\href {\doibase 10.1103/PhysRev.128.606} {\bibfield  {journal}
  {\bibinfo  {journal} {Phys. Rev.}\ }\textbf {\bibinfo {volume} {128}},\
  \bibinfo {pages} {606} (\bibinfo {year} {1962})}\BibitemShut {NoStop}%
\bibitem [{\citenamefont {Ju}\ \emph {et~al.}(2009)\citenamefont {Ju},
  \citenamefont {Cai},\ and\ \citenamefont {Guo}}]{Ju2009}%
  \BibitemOpen
  \bibfield  {author} {\bibinfo {author} {\bibfnamefont {S.}~\bibnamefont
  {Ju}}, \bibinfo {author} {\bibfnamefont {T.-Y.}\ \bibnamefont {Cai}}, \ and\
  \bibinfo {author} {\bibfnamefont {G.-Y.}\ \bibnamefont {Guo}},\ }\href
  {\doibase 10.1063/1.3146796} {\bibfield  {journal} {\bibinfo  {journal} {The
  Journal of Chemical Physics}\ }\textbf {\bibinfo {volume} {130}},\ \bibinfo
  {pages} {214708} (\bibinfo {year} {2009})},\ \Eprint
  {http://arxiv.org/abs/https://doi.org/10.1063/1.3146796}
  {https://doi.org/10.1063/1.3146796} \BibitemShut {NoStop}%
\bibitem [{\citenamefont {Young}\ and\ \citenamefont
  {Rappe}(2012)}]{Young2012}%
  \BibitemOpen
  \bibfield  {author} {\bibinfo {author} {\bibfnamefont {S.~M.}\ \bibnamefont
  {Young}}\ and\ \bibinfo {author} {\bibfnamefont {A.~M.}\ \bibnamefont
  {Rappe}},\ }\href {\doibase 10.1103/PhysRevLett.109.116601} {\bibfield
  {journal} {\bibinfo  {journal} {Phys. Rev. Lett.}\ }\textbf {\bibinfo
  {volume} {109}},\ \bibinfo {pages} {116601} (\bibinfo {year}
  {2012})}\BibitemShut {NoStop}%
\bibitem [{\citenamefont {Padmanabhan}\ \emph {et~al.}(2018)\citenamefont
  {Padmanabhan}, \citenamefont {Park}, \citenamefont {Puggioni}, \citenamefont
  {Yuan}, \citenamefont {Cao}, \citenamefont {Gasparov}, \citenamefont {Shi},
  \citenamefont {Chakhalian}, \citenamefont {Rondinelli},\ and\ \citenamefont
  {Gopalan}}]{Padmanabhan2018}%
  \BibitemOpen
  \bibfield  {author} {\bibinfo {author} {\bibfnamefont {H.}~\bibnamefont
  {Padmanabhan}}, \bibinfo {author} {\bibfnamefont {Y.}~\bibnamefont {Park}},
  \bibinfo {author} {\bibfnamefont {D.}~\bibnamefont {Puggioni}}, \bibinfo
  {author} {\bibfnamefont {Y.}~\bibnamefont {Yuan}}, \bibinfo {author}
  {\bibfnamefont {Y.}~\bibnamefont {Cao}}, \bibinfo {author} {\bibfnamefont
  {L.}~\bibnamefont {Gasparov}}, \bibinfo {author} {\bibfnamefont
  {Y.}~\bibnamefont {Shi}}, \bibinfo {author} {\bibfnamefont {J.}~\bibnamefont
  {Chakhalian}}, \bibinfo {author} {\bibfnamefont {J.~M.}\ \bibnamefont
  {Rondinelli}}, \ and\ \bibinfo {author} {\bibfnamefont {V.}~\bibnamefont
  {Gopalan}},\ }\href {\doibase 10.1063/1.5042769} {\bibfield  {journal}
  {\bibinfo  {journal} {Applied Physics Letters}\ }\textbf {\bibinfo {volume}
  {113}},\ \bibinfo {pages} {122906} (\bibinfo {year} {2018})},\ \Eprint
  {http://arxiv.org/abs/https://doi.org/10.1063/1.5042769}
  {https://doi.org/10.1063/1.5042769} \BibitemShut {NoStop}%
\bibitem [{\citenamefont {Anderson}\ and\ \citenamefont
  {Blount}(1965)}]{Anderson1965}%
  \BibitemOpen
  \bibfield  {author} {\bibinfo {author} {\bibfnamefont {P.~W.}\ \bibnamefont
  {Anderson}}\ and\ \bibinfo {author} {\bibfnamefont {E.~I.}\ \bibnamefont
  {Blount}},\ }\href {\doibase 10.1103/PhysRevLett.14.217} {\bibfield
  {journal} {\bibinfo  {journal} {Phys. Rev. Lett.}\ }\textbf {\bibinfo
  {volume} {14}},\ \bibinfo {pages} {217} (\bibinfo {year} {1965})}\BibitemShut
  {NoStop}%
\bibitem [{\citenamefont {Patankar}\ \emph {et~al.}(2018)\citenamefont
  {Patankar}, \citenamefont {Wu}, \citenamefont {Lu}, \citenamefont {Rai},
  \citenamefont {Tran}, \citenamefont {Morimoto}, \citenamefont {Parker},
  \citenamefont {Grushin}, \citenamefont {Nair}, \citenamefont {Analytis},
  \citenamefont {Moore}, \citenamefont {Orenstein},\ and\ \citenamefont
  {Torchinsky}}]{Patankar2018}%
  \BibitemOpen
  \bibfield  {author} {\bibinfo {author} {\bibfnamefont {S.}~\bibnamefont
  {Patankar}}, \bibinfo {author} {\bibfnamefont {L.}~\bibnamefont {Wu}},
  \bibinfo {author} {\bibfnamefont {B.}~\bibnamefont {Lu}}, \bibinfo {author}
  {\bibfnamefont {M.}~\bibnamefont {Rai}}, \bibinfo {author} {\bibfnamefont
  {J.~D.}\ \bibnamefont {Tran}}, \bibinfo {author} {\bibfnamefont
  {T.}~\bibnamefont {Morimoto}}, \bibinfo {author} {\bibfnamefont {D.~E.}\
  \bibnamefont {Parker}}, \bibinfo {author} {\bibfnamefont {A.~G.}\
  \bibnamefont {Grushin}}, \bibinfo {author} {\bibfnamefont {N.~L.}\
  \bibnamefont {Nair}}, \bibinfo {author} {\bibfnamefont {J.~G.}\ \bibnamefont
  {Analytis}}, \bibinfo {author} {\bibfnamefont {J.~E.}\ \bibnamefont {Moore}},
  \bibinfo {author} {\bibfnamefont {J.}~\bibnamefont {Orenstein}}, \ and\
  \bibinfo {author} {\bibfnamefont {D.~H.}\ \bibnamefont {Torchinsky}},\ }\href
  {\doibase 10.1103/PhysRevB.98.165113} {\bibfield  {journal} {\bibinfo
  {journal} {Phys. Rev. B}\ }\textbf {\bibinfo {volume} {98}},\ \bibinfo
  {pages} {165113} (\bibinfo {year} {2018})}\BibitemShut {NoStop}%
\bibitem [{\citenamefont {Lu}\ \emph {et~al.}(2022)\citenamefont {Lu},
  \citenamefont {Sayyad}, \citenamefont {S\'anchez-Mart\'{\i}nez},
  \citenamefont {Manna}, \citenamefont {Felser}, \citenamefont {Grushin},\ and\
  \citenamefont {Torchinsky}}]{Lu2022}%
  \BibitemOpen
  \bibfield  {author} {\bibinfo {author} {\bibfnamefont {B.}~\bibnamefont
  {Lu}}, \bibinfo {author} {\bibfnamefont {S.}~\bibnamefont {Sayyad}}, \bibinfo
  {author} {\bibfnamefont {M.~A.}\ \bibnamefont {S\'anchez-Mart\'{\i}nez}},
  \bibinfo {author} {\bibfnamefont {K.}~\bibnamefont {Manna}}, \bibinfo
  {author} {\bibfnamefont {C.}~\bibnamefont {Felser}}, \bibinfo {author}
  {\bibfnamefont {A.~G.}\ \bibnamefont {Grushin}}, \ and\ \bibinfo {author}
  {\bibfnamefont {D.~H.}\ \bibnamefont {Torchinsky}},\ }\href {\doibase
  10.1103/PhysRevResearch.4.L022022} {\bibfield  {journal} {\bibinfo  {journal}
  {Phys. Rev. Research}\ }\textbf {\bibinfo {volume} {4}},\ \bibinfo {pages}
  {L022022} (\bibinfo {year} {2022})}\BibitemShut {NoStop}%
\bibitem [{\citenamefont {Xu}\ \emph {et~al.}(2016{\natexlab{b}})\citenamefont
  {Xu}, \citenamefont {Dai}, \citenamefont {Zhao}, \citenamefont {Wang},
  \citenamefont {Yang}, \citenamefont {Zhang}, \citenamefont {Liu},
  \citenamefont {Xiao}, \citenamefont {Chen}, \citenamefont {Taylor},
  \citenamefont {Yarotski}, \citenamefont {Prasankumar},\ and\ \citenamefont
  {Qiu}}]{Xu2016-}%
  \BibitemOpen
  \bibfield  {author} {\bibinfo {author} {\bibfnamefont {B.}~\bibnamefont
  {Xu}}, \bibinfo {author} {\bibfnamefont {Y.~M.}\ \bibnamefont {Dai}},
  \bibinfo {author} {\bibfnamefont {L.~X.}\ \bibnamefont {Zhao}}, \bibinfo
  {author} {\bibfnamefont {K.}~\bibnamefont {Wang}}, \bibinfo {author}
  {\bibfnamefont {R.}~\bibnamefont {Yang}}, \bibinfo {author} {\bibfnamefont
  {W.}~\bibnamefont {Zhang}}, \bibinfo {author} {\bibfnamefont {J.~Y.}\
  \bibnamefont {Liu}}, \bibinfo {author} {\bibfnamefont {H.}~\bibnamefont
  {Xiao}}, \bibinfo {author} {\bibfnamefont {G.~F.}\ \bibnamefont {Chen}},
  \bibinfo {author} {\bibfnamefont {A.~J.}\ \bibnamefont {Taylor}}, \bibinfo
  {author} {\bibfnamefont {D.~A.}\ \bibnamefont {Yarotski}}, \bibinfo {author}
  {\bibfnamefont {R.~P.}\ \bibnamefont {Prasankumar}}, \ and\ \bibinfo {author}
  {\bibfnamefont {X.~G.}\ \bibnamefont {Qiu}},\ }\href {\doibase
  10.1103/PhysRevB.93.121110} {\bibfield  {journal} {\bibinfo  {journal} {Phys.
  Rev. B}\ }\textbf {\bibinfo {volume} {93}},\ \bibinfo {pages} {121110}
  (\bibinfo {year} {2016}{\natexlab{b}})}\BibitemShut {NoStop}%
\bibitem [{\citenamefont {Li}\ \emph {et~al.}(2018)\citenamefont {Li},
  \citenamefont {Jin}, \citenamefont {Tohyama}, \citenamefont {Iitaka},
  \citenamefont {Zhang},\ and\ \citenamefont {Su}}]{Li2018}%
  \BibitemOpen
  \bibfield  {author} {\bibinfo {author} {\bibfnamefont {Z.}~\bibnamefont
  {Li}}, \bibinfo {author} {\bibfnamefont {Y.-Q.}\ \bibnamefont {Jin}},
  \bibinfo {author} {\bibfnamefont {T.}~\bibnamefont {Tohyama}}, \bibinfo
  {author} {\bibfnamefont {T.}~\bibnamefont {Iitaka}}, \bibinfo {author}
  {\bibfnamefont {J.-X.}\ \bibnamefont {Zhang}}, \ and\ \bibinfo {author}
  {\bibfnamefont {H.}~\bibnamefont {Su}},\ }\href {\doibase
  10.1103/PhysRevB.97.085201} {\bibfield  {journal} {\bibinfo  {journal} {Phys.
  Rev. B}\ }\textbf {\bibinfo {volume} {97}},\ \bibinfo {pages} {085201}
  (\bibinfo {year} {2018})}\BibitemShut {NoStop}%
\bibitem [{\citenamefont {Stojchevska}\ \emph {et~al.}(2014)\citenamefont
  {Stojchevska}, \citenamefont {Vaskivskyi}, \citenamefont {Mertelj},
  \citenamefont {Kusar}, \citenamefont {Svetin}, \citenamefont {Brazovskii},\
  and\ \citenamefont {Mihailovic}}]{Stojchevska2014}%
  \BibitemOpen
  \bibfield  {author} {\bibinfo {author} {\bibfnamefont {L.}~\bibnamefont
  {Stojchevska}}, \bibinfo {author} {\bibfnamefont {I.}~\bibnamefont
  {Vaskivskyi}}, \bibinfo {author} {\bibfnamefont {T.}~\bibnamefont {Mertelj}},
  \bibinfo {author} {\bibfnamefont {P.}~\bibnamefont {Kusar}}, \bibinfo
  {author} {\bibfnamefont {D.}~\bibnamefont {Svetin}}, \bibinfo {author}
  {\bibfnamefont {S.}~\bibnamefont {Brazovskii}}, \ and\ \bibinfo {author}
  {\bibfnamefont {D.}~\bibnamefont {Mihailovic}},\ }\href {\doibase
  10.1126/science.1241591} {\bibfield  {journal} {\bibinfo  {journal}
  {Science}\ }\textbf {\bibinfo {volume} {344}},\ \bibinfo {pages} {177}
  (\bibinfo {year} {2014})},\ \Eprint
  {http://arxiv.org/abs/https://www.science.org/doi/pdf/10.1126/science.1241591}
  {https://www.science.org/doi/pdf/10.1126/science.1241591} \BibitemShut
  {NoStop}%
\bibitem [{\citenamefont {Wang}\ \emph {et~al.}(2021)\citenamefont {Wang},
  \citenamefont {Wu}, \citenamefont {Yin}, \citenamefont {Gong}, \citenamefont
  {Tu}, \citenamefont {Lin}, \citenamefont {Liu}, \citenamefont {Shi},
  \citenamefont {Zhang}, \citenamefont {Wu}, \citenamefont {Lei}, \citenamefont
  {Dong},\ and\ \citenamefont {Wang}}]{Wang2021}%
  \BibitemOpen
  \bibfield  {author} {\bibinfo {author} {\bibfnamefont {Z.~X.}\ \bibnamefont
  {Wang}}, \bibinfo {author} {\bibfnamefont {Q.}~\bibnamefont {Wu}}, \bibinfo
  {author} {\bibfnamefont {Q.~W.}\ \bibnamefont {Yin}}, \bibinfo {author}
  {\bibfnamefont {C.~S.}\ \bibnamefont {Gong}}, \bibinfo {author}
  {\bibfnamefont {Z.~J.}\ \bibnamefont {Tu}}, \bibinfo {author} {\bibfnamefont
  {T.}~\bibnamefont {Lin}}, \bibinfo {author} {\bibfnamefont {Q.~M.}\
  \bibnamefont {Liu}}, \bibinfo {author} {\bibfnamefont {L.~Y.}\ \bibnamefont
  {Shi}}, \bibinfo {author} {\bibfnamefont {S.~J.}\ \bibnamefont {Zhang}},
  \bibinfo {author} {\bibfnamefont {D.}~\bibnamefont {Wu}}, \bibinfo {author}
  {\bibfnamefont {H.~C.}\ \bibnamefont {Lei}}, \bibinfo {author} {\bibfnamefont
  {T.}~\bibnamefont {Dong}}, \ and\ \bibinfo {author} {\bibfnamefont {N.~L.}\
  \bibnamefont {Wang}},\ }\href {\doibase 10.1103/PhysRevB.104.165110}
  {\bibfield  {journal} {\bibinfo  {journal} {Phys. Rev. B}\ }\textbf {\bibinfo
  {volume} {104}},\ \bibinfo {pages} {165110} (\bibinfo {year}
  {2021})}\BibitemShut {NoStop}%
\bibitem [{\citenamefont {Bao}\ \emph {et~al.}(2022)\citenamefont {Bao},
  \citenamefont {Tang}, \citenamefont {Sun},\ and\ \citenamefont
  {Zhou}}]{Bao2022}%
  \BibitemOpen
  \bibfield  {author} {\bibinfo {author} {\bibfnamefont {C.}~\bibnamefont
  {Bao}}, \bibinfo {author} {\bibfnamefont {P.}~\bibnamefont {Tang}}, \bibinfo
  {author} {\bibfnamefont {D.}~\bibnamefont {Sun}}, \ and\ \bibinfo {author}
  {\bibfnamefont {S.}~\bibnamefont {Zhou}},\ }\href {\doibase
  10.1038/s42254-021-00388-1} {\bibfield  {journal} {\bibinfo  {journal}
  {Nature Reviews Physics}\ }\textbf {\bibinfo {volume} {4}},\ \bibinfo {pages}
  {33} (\bibinfo {year} {2022})}\BibitemShut {NoStop}%
\bibitem [{\citenamefont {Sirica}\ \emph {et~al.}(2022)\citenamefont {Sirica},
  \citenamefont {Orth}, \citenamefont {Scheurer}, \citenamefont {Dai},
  \citenamefont {Lee}, \citenamefont {Padmanabhan}, \citenamefont {Mix},
  \citenamefont {Teitelbaum}, \citenamefont {Trigo}, \citenamefont {Zhao},
  \citenamefont {Chen}, \citenamefont {Xu}, \citenamefont {Yang}, \citenamefont
  {Shen}, \citenamefont {Hu}, \citenamefont {Lee}, \citenamefont {Lin},
  \citenamefont {Cochran}, \citenamefont {Trugman}, \citenamefont {Zhu},
  \citenamefont {Hasan}, \citenamefont {Ni}, \citenamefont {Qiu}, \citenamefont
  {Taylor}, \citenamefont {Yarotski},\ and\ \citenamefont
  {Prasankumar}}]{Sirica2022}%
  \BibitemOpen
  \bibfield  {author} {\bibinfo {author} {\bibfnamefont {N.}~\bibnamefont
  {Sirica}}, \bibinfo {author} {\bibfnamefont {P.~P.}\ \bibnamefont {Orth}},
  \bibinfo {author} {\bibfnamefont {M.~S.}\ \bibnamefont {Scheurer}}, \bibinfo
  {author} {\bibfnamefont {Y.~M.}\ \bibnamefont {Dai}}, \bibinfo {author}
  {\bibfnamefont {M.-C.}\ \bibnamefont {Lee}}, \bibinfo {author} {\bibfnamefont
  {P.}~\bibnamefont {Padmanabhan}}, \bibinfo {author} {\bibfnamefont {L.~T.}\
  \bibnamefont {Mix}}, \bibinfo {author} {\bibfnamefont {S.~W.}\ \bibnamefont
  {Teitelbaum}}, \bibinfo {author} {\bibfnamefont {M.}~\bibnamefont {Trigo}},
  \bibinfo {author} {\bibfnamefont {L.~X.}\ \bibnamefont {Zhao}}, \bibinfo
  {author} {\bibfnamefont {G.~F.}\ \bibnamefont {Chen}}, \bibinfo {author}
  {\bibfnamefont {B.}~\bibnamefont {Xu}}, \bibinfo {author} {\bibfnamefont
  {R.}~\bibnamefont {Yang}}, \bibinfo {author} {\bibfnamefont {B.}~\bibnamefont
  {Shen}}, \bibinfo {author} {\bibfnamefont {C.}~\bibnamefont {Hu}}, \bibinfo
  {author} {\bibfnamefont {C.-C.}\ \bibnamefont {Lee}}, \bibinfo {author}
  {\bibfnamefont {H.}~\bibnamefont {Lin}}, \bibinfo {author} {\bibfnamefont
  {T.~A.}\ \bibnamefont {Cochran}}, \bibinfo {author} {\bibfnamefont {S.~A.}\
  \bibnamefont {Trugman}}, \bibinfo {author} {\bibfnamefont {J.-X.}\
  \bibnamefont {Zhu}}, \bibinfo {author} {\bibfnamefont {M.~Z.}\ \bibnamefont
  {Hasan}}, \bibinfo {author} {\bibfnamefont {N.}~\bibnamefont {Ni}}, \bibinfo
  {author} {\bibfnamefont {X.~G.}\ \bibnamefont {Qiu}}, \bibinfo {author}
  {\bibfnamefont {A.~J.}\ \bibnamefont {Taylor}}, \bibinfo {author}
  {\bibfnamefont {D.~A.}\ \bibnamefont {Yarotski}}, \ and\ \bibinfo {author}
  {\bibfnamefont {R.~P.}\ \bibnamefont {Prasankumar}},\ }\href {\doibase
  10.1038/s41563-021-01126-9} {\bibfield  {journal} {\bibinfo  {journal}
  {Nature Materials}\ }\textbf {\bibinfo {volume} {21}},\ \bibinfo {pages} {62}
  (\bibinfo {year} {2022})}\BibitemShut {NoStop}%
\bibitem [{\citenamefont {Morimoto}\ and\ \citenamefont
  {Nagaosa}(2016)}]{Morimoto2016}%
  \BibitemOpen
  \bibfield  {author} {\bibinfo {author} {\bibfnamefont {T.}~\bibnamefont
  {Morimoto}}\ and\ \bibinfo {author} {\bibfnamefont {N.}~\bibnamefont
  {Nagaosa}},\ }\href {\doibase 10.1126/sciadv.1501524} {\bibfield  {journal}
  {\bibinfo  {journal} {Science Advances}\ }\textbf {\bibinfo {volume} {2}},\
  \bibinfo {pages} {e1501524} (\bibinfo {year} {2016})},\ \Eprint
  {http://arxiv.org/abs/https://www.science.org/doi/pdf/10.1126/sciadv.1501524}
  {https://www.science.org/doi/pdf/10.1126/sciadv.1501524} \BibitemShut
  {NoStop}%
\bibitem [{\citenamefont {Saeta}\ \emph {et~al.}(1991)\citenamefont {Saeta},
  \citenamefont {Wang}, \citenamefont {Siegal}, \citenamefont {Bloembergen},\
  and\ \citenamefont {Mazur}}]{Saeta1991}%
  \BibitemOpen
  \bibfield  {author} {\bibinfo {author} {\bibfnamefont {P.}~\bibnamefont
  {Saeta}}, \bibinfo {author} {\bibfnamefont {J.-K.}\ \bibnamefont {Wang}},
  \bibinfo {author} {\bibfnamefont {Y.}~\bibnamefont {Siegal}}, \bibinfo
  {author} {\bibfnamefont {N.}~\bibnamefont {Bloembergen}}, \ and\ \bibinfo
  {author} {\bibfnamefont {E.}~\bibnamefont {Mazur}},\ }\href {\doibase
  10.1103/PhysRevLett.67.1023} {\bibfield  {journal} {\bibinfo  {journal}
  {Phys. Rev. Lett.}\ }\textbf {\bibinfo {volume} {67}},\ \bibinfo {pages}
  {1023} (\bibinfo {year} {1991})}\BibitemShut {NoStop}%
\bibitem [{\citenamefont {Hunsche}\ \emph {et~al.}(1995)\citenamefont
  {Hunsche}, \citenamefont {Wienecke}, \citenamefont {Dekorsy},\ and\
  \citenamefont {Kurz}}]{Hunsche1995}%
  \BibitemOpen
  \bibfield  {author} {\bibinfo {author} {\bibfnamefont {S.}~\bibnamefont
  {Hunsche}}, \bibinfo {author} {\bibfnamefont {K.}~\bibnamefont {Wienecke}},
  \bibinfo {author} {\bibfnamefont {T.}~\bibnamefont {Dekorsy}}, \ and\
  \bibinfo {author} {\bibfnamefont {H.}~\bibnamefont {Kurz}},\ }\href {\doibase
  10.1103/PhysRevLett.75.1815} {\bibfield  {journal} {\bibinfo  {journal}
  {Phys. Rev. Lett.}\ }\textbf {\bibinfo {volume} {75}},\ \bibinfo {pages}
  {1815} (\bibinfo {year} {1995})}\BibitemShut {NoStop}%
\bibitem [{\citenamefont {Hase}\ \emph {et~al.}(2002)\citenamefont {Hase},
  \citenamefont {Kitajima}, \citenamefont {Nakashima},\ and\ \citenamefont
  {Mizoguchi}}]{PhysRevLett.88.067401}%
  \BibitemOpen
  \bibfield  {author} {\bibinfo {author} {\bibfnamefont {M.}~\bibnamefont
  {Hase}}, \bibinfo {author} {\bibfnamefont {M.}~\bibnamefont {Kitajima}},
  \bibinfo {author} {\bibfnamefont {S.-i.}\ \bibnamefont {Nakashima}}, \ and\
  \bibinfo {author} {\bibfnamefont {K.}~\bibnamefont {Mizoguchi}},\ }\href
  {\doibase 10.1103/PhysRevLett.88.067401} {\bibfield  {journal} {\bibinfo
  {journal} {Phys. Rev. Lett.}\ }\textbf {\bibinfo {volume} {88}},\ \bibinfo
  {pages} {067401} (\bibinfo {year} {2002})}\BibitemShut {NoStop}%
\bibitem [{\citenamefont {Huang}\ \emph {et~al.}(2022)\citenamefont {Huang},
  \citenamefont {Yang}, \citenamefont {Teitelbaum}, \citenamefont {De~la
  Pe\~na}, \citenamefont {Sato}, \citenamefont {Chollet}, \citenamefont {Zhu},
  \citenamefont {Niedziela}, \citenamefont {Bansal}, \citenamefont {May},
  \citenamefont {Lindenberg}, \citenamefont {Delaire}, \citenamefont {Reis},\
  and\ \citenamefont {Trigo}}]{PhysRevX.12.011029}%
  \BibitemOpen
  \bibfield  {author} {\bibinfo {author} {\bibfnamefont {Y.}~\bibnamefont
  {Huang}}, \bibinfo {author} {\bibfnamefont {S.}~\bibnamefont {Yang}},
  \bibinfo {author} {\bibfnamefont {S.}~\bibnamefont {Teitelbaum}}, \bibinfo
  {author} {\bibfnamefont {G.}~\bibnamefont {De~la Pe\~na}}, \bibinfo {author}
  {\bibfnamefont {T.}~\bibnamefont {Sato}}, \bibinfo {author} {\bibfnamefont
  {M.}~\bibnamefont {Chollet}}, \bibinfo {author} {\bibfnamefont
  {D.}~\bibnamefont {Zhu}}, \bibinfo {author} {\bibfnamefont {J.~L.}\
  \bibnamefont {Niedziela}}, \bibinfo {author} {\bibfnamefont {D.}~\bibnamefont
  {Bansal}}, \bibinfo {author} {\bibfnamefont {A.~F.}\ \bibnamefont {May}},
  \bibinfo {author} {\bibfnamefont {A.~M.}\ \bibnamefont {Lindenberg}},
  \bibinfo {author} {\bibfnamefont {O.}~\bibnamefont {Delaire}}, \bibinfo
  {author} {\bibfnamefont {D.~A.}\ \bibnamefont {Reis}}, \ and\ \bibinfo
  {author} {\bibfnamefont {M.}~\bibnamefont {Trigo}},\ }\href {\doibase
  10.1103/PhysRevX.12.011029} {\bibfield  {journal} {\bibinfo  {journal} {Phys.
  Rev. X}\ }\textbf {\bibinfo {volume} {12}},\ \bibinfo {pages} {011029}
  (\bibinfo {year} {2022})}\BibitemShut {NoStop}%
\bibitem [{\citenamefont {Fritz}\ \emph {et~al.}(2007)\citenamefont {Fritz},
  \citenamefont {Reis}, \citenamefont {Adams}, \citenamefont {Akre},
  \citenamefont {Arthur}, \citenamefont {Blome}, \citenamefont {Bucksbaum},
  \citenamefont {Cavalieri}, \citenamefont {Engemann}, \citenamefont {Fahy},
  \citenamefont {Falcone}, \citenamefont {Fuoss}, \citenamefont {Gaffney},
  \citenamefont {George}, \citenamefont {Hajdu}, \citenamefont {Hertlein},
  \citenamefont {Hillyard}, \citenamefont {von Hoegen}, \citenamefont
  {Kammler}, \citenamefont {Kaspar}, \citenamefont {Kienberger}, \citenamefont
  {Krejcik}, \citenamefont {Lee}, \citenamefont {Lindenberg}, \citenamefont
  {McFarland}, \citenamefont {Meyer}, \citenamefont {Montagne}, \citenamefont
  {Murray}, \citenamefont {Nelson}, \citenamefont {Nicoul}, \citenamefont
  {Pahl}, \citenamefont {Rudati}, \citenamefont {Schlarb}, \citenamefont
  {Siddons}, \citenamefont {Sokolowski-Tinten}, \citenamefont {Tschentscher},
  \citenamefont {von~der Linde},\ and\ \citenamefont {Hastings}}]{Fritz2007}%
  \BibitemOpen
  \bibfield  {author} {\bibinfo {author} {\bibfnamefont {D.~M.}\ \bibnamefont
  {Fritz}}, \bibinfo {author} {\bibfnamefont {D.~A.}\ \bibnamefont {Reis}},
  \bibinfo {author} {\bibfnamefont {B.}~\bibnamefont {Adams}}, \bibinfo
  {author} {\bibfnamefont {R.~A.}\ \bibnamefont {Akre}}, \bibinfo {author}
  {\bibfnamefont {J.}~\bibnamefont {Arthur}}, \bibinfo {author} {\bibfnamefont
  {C.}~\bibnamefont {Blome}}, \bibinfo {author} {\bibfnamefont {P.~H.}\
  \bibnamefont {Bucksbaum}}, \bibinfo {author} {\bibfnamefont {A.~L.}\
  \bibnamefont {Cavalieri}}, \bibinfo {author} {\bibfnamefont {S.}~\bibnamefont
  {Engemann}}, \bibinfo {author} {\bibfnamefont {S.}~\bibnamefont {Fahy}},
  \bibinfo {author} {\bibfnamefont {R.~W.}\ \bibnamefont {Falcone}}, \bibinfo
  {author} {\bibfnamefont {P.~H.}\ \bibnamefont {Fuoss}}, \bibinfo {author}
  {\bibfnamefont {K.~J.}\ \bibnamefont {Gaffney}}, \bibinfo {author}
  {\bibfnamefont {M.~J.}\ \bibnamefont {George}}, \bibinfo {author}
  {\bibfnamefont {J.}~\bibnamefont {Hajdu}}, \bibinfo {author} {\bibfnamefont
  {M.~P.}\ \bibnamefont {Hertlein}}, \bibinfo {author} {\bibfnamefont {P.~B.}\
  \bibnamefont {Hillyard}}, \bibinfo {author} {\bibfnamefont {M.~H.}\
  \bibnamefont {von Hoegen}}, \bibinfo {author} {\bibfnamefont
  {M.}~\bibnamefont {Kammler}}, \bibinfo {author} {\bibfnamefont
  {J.}~\bibnamefont {Kaspar}}, \bibinfo {author} {\bibfnamefont
  {R.}~\bibnamefont {Kienberger}}, \bibinfo {author} {\bibfnamefont
  {P.}~\bibnamefont {Krejcik}}, \bibinfo {author} {\bibfnamefont {S.~H.}\
  \bibnamefont {Lee}}, \bibinfo {author} {\bibfnamefont {A.~M.}\ \bibnamefont
  {Lindenberg}}, \bibinfo {author} {\bibfnamefont {B.}~\bibnamefont
  {McFarland}}, \bibinfo {author} {\bibfnamefont {D.}~\bibnamefont {Meyer}},
  \bibinfo {author} {\bibfnamefont {T.}~\bibnamefont {Montagne}}, \bibinfo
  {author} {\bibfnamefont {E.~D.}\ \bibnamefont {Murray}}, \bibinfo {author}
  {\bibfnamefont {A.~J.}\ \bibnamefont {Nelson}}, \bibinfo {author}
  {\bibfnamefont {M.}~\bibnamefont {Nicoul}}, \bibinfo {author} {\bibfnamefont
  {R.}~\bibnamefont {Pahl}}, \bibinfo {author} {\bibfnamefont {J.}~\bibnamefont
  {Rudati}}, \bibinfo {author} {\bibfnamefont {H.}~\bibnamefont {Schlarb}},
  \bibinfo {author} {\bibfnamefont {D.~P.}\ \bibnamefont {Siddons}}, \bibinfo
  {author} {\bibfnamefont {K.}~\bibnamefont {Sokolowski-Tinten}}, \bibinfo
  {author} {\bibfnamefont {T.}~\bibnamefont {Tschentscher}}, \bibinfo {author}
  {\bibfnamefont {D.}~\bibnamefont {von~der Linde}}, \ and\ \bibinfo {author}
  {\bibfnamefont {J.~B.}\ \bibnamefont {Hastings}},\ }\href {\doibase
  10.1126/science.1135009} {\bibfield  {journal} {\bibinfo  {journal}
  {Science}\ }\textbf {\bibinfo {volume} {315}},\ \bibinfo {pages} {633}
  (\bibinfo {year} {2007})},\ \Eprint
  {http://arxiv.org/abs/https://www.science.org/doi/pdf/10.1126/science.1135009}
  {https://www.science.org/doi/pdf/10.1126/science.1135009} \BibitemShut
  {NoStop}%
\bibitem [{\citenamefont {Coulter}\ \emph {et~al.}(2018)\citenamefont
  {Coulter}, \citenamefont {Sundararaman},\ and\ \citenamefont
  {Narang}}]{Coulter2018}%
  \BibitemOpen
  \bibfield  {author} {\bibinfo {author} {\bibfnamefont {J.}~\bibnamefont
  {Coulter}}, \bibinfo {author} {\bibfnamefont {R.}~\bibnamefont
  {Sundararaman}}, \ and\ \bibinfo {author} {\bibfnamefont {P.}~\bibnamefont
  {Narang}},\ }\href {\doibase 10.1103/PhysRevB.98.115130} {\bibfield
  {journal} {\bibinfo  {journal} {Phys. Rev. B}\ }\textbf {\bibinfo {volume}
  {98}},\ \bibinfo {pages} {115130} (\bibinfo {year} {2018})}\BibitemShut
  {NoStop}%
\bibitem [{\citenamefont {Zeiger}\ \emph {et~al.}(1992)\citenamefont {Zeiger},
  \citenamefont {Vidal}, \citenamefont {Cheng}, \citenamefont {Ippen},
  \citenamefont {Dresselhaus},\ and\ \citenamefont {Dresselhaus}}]{Zeiger1992}%
  \BibitemOpen
  \bibfield  {author} {\bibinfo {author} {\bibfnamefont {H.~J.}\ \bibnamefont
  {Zeiger}}, \bibinfo {author} {\bibfnamefont {J.}~\bibnamefont {Vidal}},
  \bibinfo {author} {\bibfnamefont {T.~K.}\ \bibnamefont {Cheng}}, \bibinfo
  {author} {\bibfnamefont {E.~P.}\ \bibnamefont {Ippen}}, \bibinfo {author}
  {\bibfnamefont {G.}~\bibnamefont {Dresselhaus}}, \ and\ \bibinfo {author}
  {\bibfnamefont {M.~S.}\ \bibnamefont {Dresselhaus}},\ }\href {\doibase
  10.1103/PhysRevB.45.768} {\bibfield  {journal} {\bibinfo  {journal} {Phys.
  Rev. B}\ }\textbf {\bibinfo {volume} {45}},\ \bibinfo {pages} {768} (\bibinfo
  {year} {1992})}\BibitemShut {NoStop}%
\bibitem [{\citenamefont {Li}\ \emph {et~al.}(2022)\citenamefont {Li},
  \citenamefont {Li}, \citenamefont {Deshpande}, \citenamefont {Li},
  \citenamefont {Nair}, \citenamefont {Analytis}, \citenamefont {Silevitch},
  \citenamefont {Rosenbaum},\ and\ \citenamefont {Hsieh}}]{Li2022}%
  \BibitemOpen
  \bibfield  {author} {\bibinfo {author} {\bibfnamefont {C.}~\bibnamefont
  {Li}}, \bibinfo {author} {\bibfnamefont {X.}~\bibnamefont {Li}}, \bibinfo
  {author} {\bibfnamefont {T.}~\bibnamefont {Deshpande}}, \bibinfo {author}
  {\bibfnamefont {X.}~\bibnamefont {Li}}, \bibinfo {author} {\bibfnamefont
  {N.}~\bibnamefont {Nair}}, \bibinfo {author} {\bibfnamefont {J.~G.}\
  \bibnamefont {Analytis}}, \bibinfo {author} {\bibfnamefont {D.~M.}\
  \bibnamefont {Silevitch}}, \bibinfo {author} {\bibfnamefont {T.~F.}\
  \bibnamefont {Rosenbaum}}, \ and\ \bibinfo {author} {\bibfnamefont
  {D.}~\bibnamefont {Hsieh}},\ }\href {\doibase 10.1103/PhysRevB.106.014101}
  {\bibfield  {journal} {\bibinfo  {journal} {Phys. Rev. B}\ }\textbf {\bibinfo
  {volume} {106}},\ \bibinfo {pages} {014101} (\bibinfo {year}
  {2022})}\BibitemShut {NoStop}%
\bibitem [{\citenamefont {Hu}\ \emph {et~al.}(2022)\citenamefont {Hu},
  \citenamefont {Wu}, \citenamefont {Wang}, \citenamefont {Shi}, \citenamefont
  {Liu}, \citenamefont {Yue}, \citenamefont {Zhang}, \citenamefont {Li},
  \citenamefont {Zhou}, \citenamefont {Xu}, \citenamefont {Wu}, \citenamefont
  {Dong},\ and\ \citenamefont {Wang}}]{Hu2022}%
  \BibitemOpen
  \bibfield  {author} {\bibinfo {author} {\bibfnamefont {T.~C.}\ \bibnamefont
  {Hu}}, \bibinfo {author} {\bibfnamefont {Q.}~\bibnamefont {Wu}}, \bibinfo
  {author} {\bibfnamefont {Z.~X.}\ \bibnamefont {Wang}}, \bibinfo {author}
  {\bibfnamefont {L.~Y.}\ \bibnamefont {Shi}}, \bibinfo {author} {\bibfnamefont
  {Q.~M.}\ \bibnamefont {Liu}}, \bibinfo {author} {\bibfnamefont
  {L.}~\bibnamefont {Yue}}, \bibinfo {author} {\bibfnamefont {S.~J.}\
  \bibnamefont {Zhang}}, \bibinfo {author} {\bibfnamefont {R.~S.}\ \bibnamefont
  {Li}}, \bibinfo {author} {\bibfnamefont {X.~Y.}\ \bibnamefont {Zhou}},
  \bibinfo {author} {\bibfnamefont {S.~X.}\ \bibnamefont {Xu}}, \bibinfo
  {author} {\bibfnamefont {D.}~\bibnamefont {Wu}}, \bibinfo {author}
  {\bibfnamefont {T.}~\bibnamefont {Dong}}, \ and\ \bibinfo {author}
  {\bibfnamefont {N.~L.}\ \bibnamefont {Wang}},\ }\href {\doibase
  10.1103/PhysRevB.105.075113} {\bibfield  {journal} {\bibinfo  {journal}
  {Phys. Rev. B}\ }\textbf {\bibinfo {volume} {105}},\ \bibinfo {pages}
  {075113} (\bibinfo {year} {2022})}\BibitemShut {NoStop}%
\end{thebibliography}%

\end{document}